\documentclass{article}
\usepackage{graphicx} 
\usepackage{amsmath, amssymb}
\usepackage{amsmath, amsfonts}
\usepackage[dvipsnames]{xcolor} 
\usepackage{algorithm}
\usepackage{algpseudocode}
\usepackage{amsmath}
\usepackage{multirow}
\usepackage{subfig}
\usepackage{makecell} 
\usepackage{hyperref}
\usepackage{tikz}

\usepackage[markup=underlined]{changes}

 \usetikzlibrary{arrows.meta,positioning,fit,calc,shapes.multipart}

\usepackage{mathrsfs}%
\usepackage{authblk}
\usepackage{fontawesome5}
\usepackage[pagewise]{lineno}

\title{Deep Reinforcement Learning for Optimizing
Angle Selection and Dose Allocation in CT
Reconstruction}
\author[1,2]{%
  Tianyuan Wang\thanks{Corresponding author: \href{mailto:tianyuan.wang@cwi.nl}{tianyuan.wang@cwi.nl}}%
  \,\href{mailto:tianyuan.wang@cwi.nl}{\faEnvelope}%
}

\author[2]{Daniël M. Pelt \texorpdfstring{%
    \textsuperscript{\,\href{mailto:d.m.pelt@liacs.leidenuniv.nl}{\faEnvelope}}
  }{}%
}

\author[1]{Felix Lucka \texorpdfstring{%
    \textsuperscript{\,\href{mailto:Felix.Lucka@cwi.nl}{\faEnvelope}}
  }{}%
}

\author[1,3]{Tristan van Leeuwen\texorpdfstring{%
    \textsuperscript{\,\href{mailto:T.van.Leeuwen@cwi.nl}{\faEnvelope}}
  }{}%
}

\author[1,2]{Kees Joost Batenburg \texorpdfstring{%
    \textsuperscript{\,\href{mailto:k.j.batenburg@liacs.leidenuniv.nl}{\faEnvelope}}
  }{}%
}

\affil[1]{Centrum Wiskunde \& Informatica, Science Park 123, Amsterdam, 1098 XG, The Netherlands}
\affil[2]{Leiden Institute of Advanced Computer Science, Leiden Universiteit, Leiden, The Netherlands}
\affil[3]{Mathematics Institute, Utrecht University, Campus-Boulevard 30, Utrecht, 3584 CD, The Netherlands}
\title{Deep Reinforcement Learning for Optimizing Angle Selection and Dose Allocation in CT Reconstruction}

\begin{document}

\maketitle

\begin{abstract}
Traditional X-ray computed tomography (CT) scanning strategies typically select projection angles uniformly and allocate dose equally. 
In practice, however, CT scans often need to be fast, radiation-efficient, and adaptive. 
Sparse-view tomography addresses these requirements by reducing both the number of angles and the total dose budget. 
Under such constraints, angle selection and dose allocation should be information-driven, with more dose assigned to informative directions. 
To this end, we propose a dose-aware acquisition and reconstruction framework
that combines a PWLS–PnP reconstruction backbone with an RL-based strategy for
adaptive angle selection, explicitly accounting for angle-dependent photon
statistics. Numerical experiments show that the proposed approach improves overall reconstruction quality and enhances defect detectability compared with conventional strategies, particularly when only a small number of projections or a constrained dose budget is available.  


\end{abstract}

\newpage

\section{Introduction}
X-ray Computed Tomography (CT) is a widely used non-destructive imaging modality that reconstructs the internal structure of objects from external projections. However, CT inherently faces trade-offs among image quality, scan time, and radiation exposure—particularly in settings requiring rapid inspection or involving dose-sensitive subjects.

In medical CT, \emph{dose optimization} is critical to minimize patient radiation exposure. As reviewed in~\cite{mccollough2006ct}, conventional approaches such as tube current modulation and automatic exposure control adjust exposure according to patient size and anatomy. Although effective, these strategies remain largely heuristic and lack task-specific adaptivity. In \emph{industrial CT}, the objective is often to concentrate measurements on regions with high uncertainty or suspected defects while avoiding redundant projections. Similar considerations arise in computational microscopy and electron or optical tomography, where optimizing illumination dose is essential to prevent specimen damage. The object’s position or orientation may differ between scans due to manipulation or changes in setup. These factors motivate the development of \emph{adaptive scanning strategies} that jointly optimize projection angle selection and dose allocation under a fixed dose budget. Depending on the imaging task, the total dose can be distributed over many low-intensity views or concentrated on fewer, more informative angles, since not all measurements contribute equally to reconstruction quality~\cite{kazantsev1991information}. 

Several works investigate adaptive scanning strategies that focus solely on sequential angle selection, where projection views are chosen based on information acquired during the scan~\cite{batenburg2013dynamic, elata2025adaptive, barbano2022bayesian, wang2024sequential}. The problem of selecting informative projection angles has often been formulated as a \emph{bi-level optimization} problem~\cite{ruthotto2018optimal}, where the lower level performs image reconstruction, and the upper level minimizes reconstruction error. This approach is high-dimensional and non-convex, as both levels depend on the design parameters (angles). Alternatively, \emph{Bayesian experimental design}~\cite{lindley1972bayesian, ryan2016fully, rainforth2024modern} seeks to maximize expected information gain by comparing prior and posterior distributions or reconstruction accuracy. Extensions of this framework have incorporated prior knowledge to guide angle selection and improve acquisition efficiency~\cite{batenburg2011bounds, batenburg2013dynamic, dabravolski2014dynamic, helin2022edge, barbano2022bayesian, elata2025adaptive}. However, extending these methods to include \emph{dose allocation} further increases dimensionality and computational cost. In practical applications, imaging often requires adaptive, online decision-making to accommodate variations in object geometry or motion. Traditional Bayesian approaches, which rely on repeated evaluation of expected information gain, are computationally intensive and unsuitable for real-time control. Moreover, the \emph{joint optimization of angle and dose} remains largely unexplored, despite its strong impact on reconstruction quality.

Recent works have explored \emph{Reinforcement Learning (RL)} for sequential CT scanning~\cite{shen2022learning, wang2024sequential, wang2024task, wang2025dynamic}. Policy-based RL methods learn scanning strategies directly from data, eliminating repeated optimization during acquisition and enabling adaptive decision-making based on intermediate reconstructions. Once trained, these policies can be applied efficiently in online scanning scenarios.

From a reconstruction perspective, the \emph{incident photon count}—that is, the local dose—directly affects measurement noise according to the Beer–Lambert law. When dose allocation varies across angles, high-dose projections yield lower noise, while low-dose projections are more uncertain. This heterogeneity calls for reconstruction algorithms that explicitly account for dose-dependent noise. \emph{Penalized Weighted Least Squares (PWLS)} methods~\cite{wang2006penalized, kazantsev2017novel} address noise in low-dose CT but typically assume uniform dose across angles. Some learning-based approaches incorporate global dose levels~\cite{xia2021ct}, yet they do not handle unequal per-angle dose distributions.

In this work, we address this gap by deriving a \emph{dose-aware PWLS} model derived from a second-order Taylor approximation of the Poisson log-likelihood. 
While conventional PWLS formulations generally assume uniform or fixed doses across views, our adaptive setting requires handling intentionally non-uniform, per-angle dose allocations. 
To accommodate this, we introduce an angle-dependent weighting matrix that explicitly reflects the dose assigned to each view by the adaptive policy.

\subsection*{Contributions}
Building on recent advances in adaptive imaging, this work introduces a unified RL-based framework for joint angle and dose optimization in X-ray CT. The main contributions are:
\begin{itemize}
    \item \textbf{Joint adaptive scanning framework:} An RL-based system that jointly learns policies for selecting projection angles and allocating dose, optimizing task-specific metrics such as reconstruction quality and defect detectability.
    \item \textbf{Dose-aware reconstruction algorithm:} 
    We adapt the conventional PWLS formulation to incorporate \emph{angle-dependent weights} consistent with the non-uniform dose distributions produced by the RL policy. 
    This dose-weighted PWLS term is combined with a Total Variation (TV) prior within a Plug-and-Play (PnP) framework to provide robust reconstruction in settings where per-angle dose is highly heterogeneous. 
\end{itemize}

The remainder of this paper is organized as follows. Section~\ref{sec:background} reviews background on CT reconstruction and sequential optimal experimental design using RL. Section~\ref{sec:methods} presents the proposed reconstruction and scanning framework. Section~\ref{sec:results} reports experimental results, and Sections~\ref{sec:discussion}--\ref{sec:conclusion} provide discussion and conclusions.

\section{Background}
\label{sec:background}




\subsection{CT Reconstruction}

In X-ray CT, the objective is to reconstruct an unknown image \(\bar{\boldsymbol{x}} \in \mathbb{R}^n\), representing the spatial distribution of attenuation coefficients within an object, from a set of external measurements. 
These measurements, denoted \(\boldsymbol{y}(\boldsymbol{\theta})\), are acquired at a sequence of design parameters \(\boldsymbol{\theta} = \{\theta_1, \dots, \theta_M\}\), typically projection angles. 
The image formation process is described by a forward operator \(\boldsymbol{A}(\boldsymbol{\theta})\), which discretizes the Radon transform under the specified acquisition geometry~\cite{hansen2021computed}.

To account for measurement uncertainty, we adopt a statistical observation model in which the measurements are treated as random variables:
\begin{equation}
    \boldsymbol{y}(\boldsymbol{\theta}) \sim \pi_{\text{data}}(\boldsymbol{y} \mid \bar{\boldsymbol{x}}; \boldsymbol{\theta}),
    \label{eq:forwardmodel}
\end{equation}
where \(\pi_{\text{data}}\) denotes the data likelihood. 

\paragraph{Beer–Lambert law.}
X-ray attenuation follows the Beer–Lambert law \cite{hansen2021computed}, which relates transmitted intensity to the cumulative attenuation along an X-ray path. 
Let \(I_0(\theta_j)\) denote the incident photon flux (dose) at angle \(\theta_j\), and \(I_i(\theta_j)\) the transmitted intensity measured at detector bin \(i\). 
Then the attenuation model is
\begin{equation}
    I_i(\theta_j) = I_0(\theta_j) \exp\!\left(-[\boldsymbol{A}(\theta_j)\bar{\boldsymbol{x}}]_i\right),
\label{eq:beerlambertlaw}
\end{equation}
where \([\boldsymbol{A}(\theta_j)\bar{\boldsymbol{x}}]_i\) represents the line integral of \(\bar{\boldsymbol{x}}\) along ray \(i\) at angle \(\theta_j\). 
The corresponding noise-free measurement is
\begin{equation}
    y_i(\theta_j) = -\log\!\left(\frac{I_i(\theta_j)}{I_0(\theta_j)}\right) = [\boldsymbol{A}(\theta_j)\bar{\boldsymbol{x}}]_i.
\end{equation}
In practice, the observed measurements \(\boldsymbol{y}(\theta_j)\) deviate from this ideal due to stochastic noise governed by the dose level \(I_0(\theta_j)\).

\paragraph{Bayesian inverse formulation.}
Reconstructing \(\bar{\boldsymbol{x}}\) from noisy measurements is an ill-posed inverse problem. 
Within a Bayesian framework, the posterior distribution combines the measurement model with prior information:
\begin{equation}
\begin{aligned}
    \widehat{\boldsymbol{x}}(\boldsymbol{\theta}) 
    &= \arg\max_{\boldsymbol{x}} \, \pi_{\text{post}}(\boldsymbol{x} \mid \boldsymbol{y}; \boldsymbol{\theta}) \\
    &= \arg\max_{\boldsymbol{x}} \, \frac{\pi_{\text{data}}(\boldsymbol{y} \mid \boldsymbol{x}; \boldsymbol{\theta}) \cdot \pi_{\text{prior}}(\boldsymbol{x})}{\pi(\boldsymbol{y})},
    \label{eq:inverseMAP}
\end{aligned}
\end{equation}
where \(\pi_{\text{prior}}(\boldsymbol{x})\) encodes assumptions about the image, such as smoothness or sparsity. 
The maximum a posteriori (MAP) estimate thus balances data fidelity and prior knowledge to yield a stable reconstruction.

\subsection{Sequential Optimal Experimental Design via Reinforcement Learning}
\label{sec:sequential-oed}

Reinforcement learning (RL) provides a natural framework for solving sequential optimal experimental design (OED) problems in CT imaging by modeling the acquisition process as a \emph{Partially Observable Markov Decision Process (POMDP)}~\cite{shen2023bayesian}. This perspective has been effectively applied in prior work~\cite{wang2024sequential, wang2024task, wang2025dynamic}, where the agent interacts with the CT environment to iteratively select measurements and refine the reconstruction.

The interaction starts from an initial estimate \(\widehat{\boldsymbol{x}}_1\), typically a zero image. At each time step \(k\), the agent selects an acquisition angle \(\theta_k\) based on the current reconstruction \(\widehat{\boldsymbol{x}}_k\). A noisy measurement \(\boldsymbol{y}_k\) is then acquired using the forward model in Equation~\eqref{eq:forwardmodel}, and a new reconstruction \(\widehat{\boldsymbol{x}}_{k+1}\) is computed based on all measurements up to step \(k\). A reward \(r_k\) is subsequently calculated by comparing \(\widehat{\boldsymbol{x}}_{k+1}\) to the ground truth image.

The agent learns a parameterized policy \(\pi_{\text{a}}(\theta \mid \widehat{\boldsymbol{x}}_k; \boldsymbol{w}_{a})\), which maps the current reconstruction to a probability distribution over candidate angles. After \(M\) steps, the acquisition process produces a trajectory \(\boldsymbol{\tau}\) defined as:
\[
\boldsymbol{\tau} = \left\{ \widehat{\boldsymbol{x}}_1, (\theta_1, \widehat{\boldsymbol{x}}_2, r_1), \dots, (\theta_M, \widehat{\boldsymbol{x}}_{M+1}, r_M) \right\}.
\]
This trajectory follows a distribution governed by:
\begin{equation}
    \pi_{\text{chain}}(\boldsymbol{\tau}; \boldsymbol{w}_a) = \prod_{k=1}^{M} \pi_{\text{a}}(\theta_k \mid \widehat{\boldsymbol{x}}_k; \boldsymbol{w}_a) \cdot \pi_{t}(\widehat{\boldsymbol{x}}_{k+1} \mid \widehat{\boldsymbol{x}}_k, \theta_k),
    \label{eq:ChainStandard}
\end{equation}
where \(\pi_{t}(\widehat{\boldsymbol{x}}_{k+1} \mid \widehat{\boldsymbol{x}}_k, \theta_k)\) represents the transition model capturing the effect of measurement and reconstruction.

The learning objective is to maximize the expected cumulative reward, where future rewards may be discounted using a factor \(\gamma \in [0,1]\) to prioritize immediate outcomes. The state-value function under policy \(\pi_{\text{a}}\) is defined as:
\begin{equation}
    V^{\pi_{\text{a}}}(\widehat{\boldsymbol{x}}_k) = \mathbb{E}_{\boldsymbol{\tau} \sim \pi_{\text{chain}}} \left[ \sum_{t=k}^{M} \gamma^{\,t-k} r_t \,\middle|\, \widehat{\boldsymbol{x}}_k
    \right],
    \label{eq:ValueStandard}
\end{equation}
where \(\gamma^{t-k}\) discounts the reward obtained at time step \(k\) relative to
the current step \(k\). This formulation captures the preference for receiving rewards earlier rather than later.

The overall objective in RL is to maximize the expected return starting from the fixed initial zero image \(\widehat{\boldsymbol{x}}_{1}\)::
\begin{equation}
    J(\boldsymbol{w}_{a}) = V^{\pi_{\text{a}}}(\widehat{\boldsymbol{x}}_1),
    \label{eq:ObjectiveStandard}
\end{equation}
where optimization is performed over the policy parameters \(\boldsymbol{w}_a\).

\subsection{Policy Gradient Methods}
\label{sec:policy-gradient}

Policy gradient methods aim to optimize the expected return \(J(\boldsymbol{w}_a)\) by computing gradients with respect to the policy parameters \(\boldsymbol{w}_a\). A popular approach is the \emph{actor-critic} framework, where the \textit{actor} represents the policy, and the \textit{critic} estimates the state-value function to guide policy improvement.

To reduce variance and improve learning efficiency, policy updates are often scaled by an \emph{advantage estimate} \(\widehat{A}_k\), which measures how much better (or worse) an action is compared to the critic’s baseline value estimate \(V^{\pi_{\text{a}}}(\widehat{\boldsymbol{x}}_k)\). A simple and widely used choice is the one-step Temporal-Difference (TD(0)) estimator \cite{sutton2018reinforcement}:
\begin{equation}
    \widehat{A}_k = r_k + \gamma V^{\pi_{\text{a}}}(\widehat{\boldsymbol{x}}_{k+1}) - V^{\pi_{\text{a}}}(\widehat{\boldsymbol{x}}_k),
\end{equation}
where \(\gamma\) is the discount factor. More general estimators, such as Generalized Advantage Estimation (GAE)~\cite{schulman2015high}, can further reduce variance by incorporating multi-step returns.

The policy gradient can then be approximated from sampled trajectories as:
\begin{equation}
\begin{aligned}
    \nabla_{\boldsymbol{w}_{a}} J(\boldsymbol{w}_{a})
    &\approx \frac{1}{N} \sum_{n=1}^{N} \Biggl(\sum_{k=1}^{M} 
    \nabla_{\boldsymbol{w}_{a}} \log \pi_{\text{a}}\!\left(\theta_k \mid \widehat{\boldsymbol{x}}_k; \boldsymbol{w}_{a}\right) 
    \cdot \widehat{A}_k\Biggr),
\end{aligned}
\label{eq:policy_gradient_advantage}
\end{equation}
where \(N\) is the number of sampled trajectories and \(M\) is the number of timesteps per trajectory.

\section{Methods}
\label{sec:methods}

This section introduces the proposed framework, which combines a dose-aware reconstruction algorithm with prior information to handle unequal dose allocation, and a RL approach to address the adaptive CT scanning problem.



\subsection{Reconstruction with Unequal Dose Allocation}

Building on the Beer–Lambert formulation in Section~\ref{sec:background}, we now incorporate measurement noise arising from photon statistics. 
Since photon detection follows a Poisson distribution, the measured photon count at detector bin $i$ for angle $\theta_j$ can be modeled as (according to Equation~\eqref{eq:beerlambertlaw})
\begin{equation}
    I_i(\theta_j) \sim \mathrm{Poisson}\!\left( I_0(\theta_j) \, \exp\!\left(-[\boldsymbol{A}(\theta_j)\bar{\boldsymbol{x}}]_i \right) \right).
\end{equation}
where $\boldsymbol{I}_0(\theta_j) = [I_{0,1}(\theta_j),\ldots,I_{0,n_{\text{det}}}(\theta_j)]^T$ denotes the incident photon flux at all detector bins.  
(For the experiments in this paper, we consider a spatially uniform incident dose so that $\boldsymbol{I}_0(\theta_j) = I_0(\theta_j)\boldsymbol{1}$.)

Normalizing by $\boldsymbol{I}_0(\theta_j)$ and applying a negative logarithm yields the noisy log-transformed measurements:
\begin{equation}
   \boldsymbol{y}(\theta_j)
    = -\log\!\left(\tfrac{\boldsymbol{I}(\theta_j)}{\boldsymbol{I}_0(\theta_j)}\right).
\end{equation}

From the Bayesian inverse formulation in Equation~\eqref{eq:inverseMAP}, the negative log-likelihood (up to constants) gives the dose-aware reconstruction objective:
\begin{equation}
\begin{aligned}
    \widehat{\boldsymbol{x}}(\boldsymbol{\theta}) \propto \arg\min_{\boldsymbol{x}} \Big\{ &
    \sum_{j} \boldsymbol{I}_0(\theta_j)^{T}
        \exp\!\left(-\boldsymbol{A}(\theta_j)\boldsymbol{x}\right) \\
    &+ \sum_{j} \left[\boldsymbol{I}_0(\theta_j)
        \odot \exp\!\big(-\boldsymbol{y}(\theta_j)\big)\right]^T
        \boldsymbol{A}(\theta_j)\boldsymbol{x}
    - \log \pi_{\text{prior}}(\boldsymbol{x})
    \Big\}.
\end{aligned}
\label{eq:inversePoisson_theta_unequal}
\end{equation}

This formulation explicitly incorporates unequal per-angle dose levels $I_0(\theta_j)$, which directly affect both the noise characteristics of the measurements and their contribution to the reconstruction objective.

\paragraph*{PWLS formulation.}
Direct minimization of Equation~\eqref{eq:inversePoisson_theta_unequal} is computationally demanding.  
Using a second-order Taylor expansion around the measured log-data \cite{hansen2021computed} yields a Penalized Weighted Least-Squares (PWLS) problem \cite{fessler1994penalized,wang2006penalized,hansen2021computed}:
\begin{equation}
    \widehat{\boldsymbol{x}}(\boldsymbol{\theta}) \propto \arg\min_{\boldsymbol{x}} \left\{
        \tfrac{1}{2}\!\left( \boldsymbol{y}(\boldsymbol{\theta}) - \boldsymbol{A}(\boldsymbol{\theta}) \boldsymbol{x} \right)^{\!T}
        \boldsymbol{W}(\boldsymbol{\theta})
        \left( \boldsymbol{y}(\boldsymbol{\theta}) - \boldsymbol{A}(\boldsymbol{\theta}) \boldsymbol{x} \right)
        - \log \pi_{\text{prior}}(\boldsymbol{x})
    \right\},
\end{equation}
with diagonal weight matrix
\begin{equation}
    \boldsymbol{W}(\boldsymbol{\theta}) = \mathrm{diag}\!\Big(
        \boldsymbol{I}_0(\boldsymbol{\theta}) \odot
        \exp\!\big(-\boldsymbol{y}(\boldsymbol{\theta})\big)
    \Big).
\end{equation}
Here, $\boldsymbol{I}_0(\boldsymbol{\theta})$ expands the per-angle dose allocation $I_0(\theta_j)$ to all bins.  
In contrast to conventional PWLS formulations \cite{wang2006penalized} that assume a uniform incident dose across views, this model explicitly accounts for angle-dependent dose allocation.
This \emph{dose-aware} weighting provides a principled mechanism to assign higher statistical confidence to high-dose projections and lower confidence to noisy low-dose measurements, effectively capturing the signal-dependent variance characteristic of Poisson noise in X-ray CT.



\paragraph{Plug-and-Play priors}
In particular when using low radiation dosages distributed over few projection angles, minimizing the PWLS objective directly is ill-posed and leads to very noisy reconstructions ridden with sparse-angle artifacts. To regularize the inversion in the most flexible way, we describe here how to employ the Plug-and-Play (PnP) framework~\cite{venkatakrishnan2013plug}. PnP decouples the reconstruction into two alternating steps: a gradient step for the PWLS objective followed by the application of an image denoiser. 
To illustrate this, we implemented denoising with a Total Variation (TV) energy \cite{rudin1992nonlinear,chambolle2004algorithm}, denoted by $\operatorname{prox}_{\tau\,\mathrm{TV}}(\cdot)$. TV penalizes the $\ell_1$-norm of image gradients and is used to suppresses noise in homogeneous regions while preserving sharp edges. While this choice means that the PnP framework reduces to a proximal gradient descent with an explicit image prior $\pi_{\text{prior}}(\boldsymbol{x})$, using, e.g., a trained DNN would correspond to using an implicit prior.


\begin{algorithm}[H]
\caption{Dose-Aware PnP–PWLS Reconstruction (Gradient + TV Prior)}
\label{alg:dose_aware_pnp}
\begin{algorithmic}[1]
\Require 
Forward operator $\boldsymbol{A}(\boldsymbol{\theta})$;  
measured log-projections $\boldsymbol{y}(\boldsymbol{\theta})$;  
dose allocation $\boldsymbol{d}(\boldsymbol{\theta})$;  
number of iterations $T$;  
initial estimate $\boldsymbol{x}_0$;  
PnP denoiser strength $\tau>0$ (TV parameter);  
positivity flag.  
\Ensure Reconstructed image $\boldsymbol{x}_T$.

\vspace{0.25em}
\State $\boldsymbol{x} \gets \boldsymbol{x}_0$
\State Normalize dose: $\boldsymbol{d}(\boldsymbol{\theta}) \gets \boldsymbol{d}(\boldsymbol{\theta}) / \mathrm{mean}(\boldsymbol{d}(\boldsymbol{\theta}))$
\State Dose–Poisson weights: $\boldsymbol{R}(\boldsymbol{\theta}) \gets \boldsymbol{d}(\boldsymbol{\theta}) \odot \exp\!\big(-\boldsymbol{y}(\boldsymbol{\theta})\big)$
\State Define weighting operator: $\mathcal{W}(\cdot): \boldsymbol{u} \mapsto \boldsymbol{R}(\boldsymbol{\theta}) \odot \boldsymbol{u}$
\State Estimate Lipschitz constant $L$ via Appendix~\ref{sec:power-iteration}; set step size $\alpha \gets h/L$ with $h \in [0,2]$

\For{$k=1$ \textbf{to} $T$} \Comment{Data fidelity + PnP prior}
    \State $\boldsymbol{y}_{\text{pred}} \gets \boldsymbol{A}(\boldsymbol{\theta})\,\boldsymbol{x}$ \hfill \Comment{Forward projection}
    \State $\boldsymbol{r} \gets \mathcal{W}\!\left(\boldsymbol{y}_{\text{pred}} - \boldsymbol{y}(\boldsymbol{\theta})\right)$ \hfill \Comment{Dose-aware residual}
    \State $\boldsymbol{g} \gets \boldsymbol{A}(\boldsymbol{\theta})^\top \boldsymbol{r}$ \hfill \Comment{Backprojection (gradient of data term)}
    \State $\boldsymbol{x} \gets \boldsymbol{x} - \alpha\,\boldsymbol{g}$ \hfill \Comment{Gradient step (data fidelity)}
    \State $\boldsymbol{x} \gets \operatorname{prox}_{\tau\,\mathrm{TV}}(\boldsymbol{x})$ \hfill \Comment{PnP denoiser: TV proximal operator (e.g.\ Chambolle’s algorithm)}
    \If{positivity flag} \State $\boldsymbol{x} \gets \max(\boldsymbol{x},0)$ \EndIf
\EndFor
\State \Return $\boldsymbol{x}$
\end{algorithmic}
\label{alg:dose_weighted_PnP_reconstruction}
\end{algorithm}

\noindent\textit{Explanation.}  
Algorithm~\ref{alg:dose_weighted_PnP_reconstruction} performs gradient descent on the PWLS objective with explicit modeling of angle-dependent dose allocation.  
At each iteration, the current reconstruction is forward-projected and compared with the measured log-projections to form a residual.  
This residual is reweighted by a Poisson–dose factor, which increases the influence of high-dose views and suppresses noisy low-dose views.  
The weighted residual is then backprojected to compute the gradient, and the step size is normalized by the estimated Lipschitz constant (Appendix~\ref{sec:power-iteration}) to ensure stable convergence.  
After the gradient update, a TV denoiser $\operatorname{prox}_{\tau\,\mathrm{TV}}(\cdot)$ is applied in the PnP framework, providing noise suppression while preserving edges.  
Finally, a positivity constraint can be optionally enforced to maintain physically valid attenuation values.  

Compared with standard iterative reconstruction, this dose-aware PnP scheme explicitly accounts for non-uniform photon allocation across projection angles, yielding statistically consistent updates under unequal-dose settings while benefiting from edge-preserving regularization.

\subsection{Proximal Policy Optimization for CT Environment}

To stably optimize the objective \(J(\boldsymbol{w}_{a})\) in Equation (\ref{eq:ObjectiveStandard}), Proximal Policy Optimization (PPO)~\cite{schulman2017proximal} is adopted as the RL framework. 
The vanilla policy-gradient method in Equation (\ref{eq:policy_gradient_advantage}) updates the policy parameters by directly
ascending an unbiased stochastic estimate of \(\nabla_{\boldsymbol{w}_a} J(\boldsymbol{w}_a)\),
computed from sampled trajectories, without explicitly constraining the size
of policy updates.
PPO replaces this unconstrained update with a clipped surrogate objective that
approximates the same gradient while preventing excessively large policy changes, thereby improving training stability:
\begin{equation}
    Loss^{\text{actor}}(\boldsymbol{w}_a) =
    \frac{1}{N}\sum_{n=1}^N \sum_{k=1}^M
    \min\!\Big(
        e_k(\boldsymbol{w}_a)\,\widehat{A}_k,\;
        c_{1-\epsilon}^{\,1+\epsilon}\!\big(e_k(\boldsymbol{w}_a)\big)\,\widehat{A}_k
    \Big),
\label{eq:ppo_loss}
\end{equation}

where $e_k(\boldsymbol{w}_a) = \tfrac{\pi_{a}(\theta_k \mid \widehat{\boldsymbol{x}}_k; \boldsymbol{w}_a)}{\pi_{a}^{\text{old}}(\theta_k \mid \widehat{\boldsymbol{x}}_k)}$ is the probability ratio between new and old policies, $\widehat{A}_k$ is the estimated advantage, and $\epsilon > 0$ is a clipping parameter. Here, $c_{1-\epsilon}^{\,1+\epsilon}(\cdot)$ clips the probability ratio 
$e_k(\boldsymbol{w}_a)$ to the interval $[1-\epsilon,\,1+\epsilon]$, 
ensuring stable and conservative policy updates.
The objective is maximized during training to update the actor parameters $\boldsymbol{w}_a$.

The critic is trained using a mean squared error (MSE) loss with respect to the Monte Carlo return $G_k$ (Formally, \(G_k = \sum_{t=k}^{M} \gamma^{\,t-k} r_t\).):
\begin{equation}
    Loss^{\text{critic}}(\boldsymbol{w}_v) =
    \frac{1}{N}\sum_{n=1}^N \sum_{k=1}^M
    \big(G_k - V(\widehat{\boldsymbol{x}}_k;\boldsymbol{w}_v)\big)^2,
\label{eq:ppo_critic}
\end{equation}

\paragraph{Reward Definition.}
The reward function depends on the objective of the reconstruction task. 
For general image quality assessment, we use the Peak Signal-to-Noise Ratio (PSNR), computed over a region of interest (RoI)~$\Omega$:
\begin{equation}
    \mathrm{PSNR}_{\Omega}(\bar{\boldsymbol{x}}, \widehat{\boldsymbol{x}}) =
    10 \log_{10}\!\left(
        \frac{H^2}{\tfrac{1}{|\Omega|}\sum_{i \in \Omega} (\bar{x}_i - \widehat{x}_i)^2}
    \right),
    \label{eq:PSNR}
\end{equation}
where $H$ denotes the dynamic range of the image.

For defect-focused tasks, the reward can instead be based on the Contrast-to-Noise Ratio (CNR):
\begin{equation}
    \mathrm{CNR}_{\mathcal{D}, \mathcal{B}}(\bar{\boldsymbol{x}}, \widehat{\boldsymbol{x}}) =
    \frac{\mu_{\mathcal{D}} - \mu_{\mathcal{B}}}{\sigma_{\mathcal{B}}},
    \label{eq:CNR}
\end{equation}
where $\mu_{\mathcal{D}}$ is the mean intensity within the defect mask $\mathcal{D}$, and $\mu_{\mathcal{B}}$ and $\sigma_{\mathcal{B}}$ denote the mean and standard deviation in a background region $\mathcal{B}$.

\paragraph{Training Procedure.}
The RL agent interacts with the CT environment under a fixed dose budget $B$, which is spent in base quanta $\Delta b$ per projection. 
Dose allocation is modeled implicitly by repeating angle selections: if an angle is chosen $z$ times, the allocated dose is $z \Delta b$. 
The episode terminates after $L=\lfloor B/\Delta b\rfloor$ steps,
corresponding to exhaustion of the total dose budget $B$.
The policy maps the current reconstruction $\widehat{\boldsymbol{x}}_k$ to a probability distribution over candidate projection angles, thereby jointly determining angle selection and dose allocation. The overall training loop is summarized in Algorithm~\ref{alg:ppo-ct}. A trajectory buffer stores observations, actions, rewards, and log-probabilities of the old policy. After each rollout, advantages and returns are computed, and the policy and value networks are updated by Equations ~(\ref{eq:ppo_loss})–(\ref{eq:ppo_critic}). Implementation details including advantage estimation, entropy regularization, and minibatch updates are provided in Appendix~\ref{appendix:ppo}.
\begin{algorithm}[H]
\caption{PPO Training in a CT Environment with a Dose Budget}
\label{alg:ppo-ct}
\begin{algorithmic}[1]
\Require Environment $\mathcal{E}$, policy $\pi_a(\cdot;\boldsymbol{w}_a)$, value function $V(\cdot;\boldsymbol{w}_v)$, total dose budget $B$, fixed dose per step $\Delta b$,
number of trajectories per iteration $n$
\State Set trajectory length $L \gets \left\lfloor \frac{B}{\Delta b} \right\rfloor$
\State Set horizon $T \gets n \cdot L$ \Comment{total steps per PPO iteration}

\For{each iteration}
    \State Initialize empty trajectory buffer $\mathcal{T}$
    \For{$i=1$ \textbf{to} $n$} \Comment{collect $n$ trajectories}
        \State Reset environment: $\widehat{\boldsymbol{x}}_{1} (\text{zero matrix}) \gets \mathcal{E}.\text{reset}()$
        \For{$k=1$ \textbf{to} $L$}
            \State Sample angle $\theta_k \sim \pi_a(\cdot \mid \widehat{\boldsymbol{x}}_{k};\boldsymbol{w}_a)$
            \State $\log\pi_k \gets \log \pi_a(\theta_k \mid \widehat{\boldsymbol{x}}_{k};\boldsymbol{w}_a)$,\quad $v_k \gets V(\widehat{\boldsymbol{x}}_{k};\boldsymbol{w}_v)$
            \State Apply action: $(\widehat{\boldsymbol{x}}_{k+1}, r_k) \gets \mathcal{E}.\text{step}(\theta_k)$ 
            \State $d_k \gets \mathbb{I}[k = L]$
            \State Store $(\widehat{\boldsymbol{x}}_{k}, \theta_k, r_k, \widehat{\boldsymbol{x}}_{k+1}, \log\pi_k, v_k, d_k)$ in $\mathcal{T}$
        \EndFor
    \EndFor
    \State Compute advantages $\widehat{A}_k$ and returns $G_k$ from $\mathcal{T}$ (Appendix~\ref{appendix:ppo})
    \State Normalize advantages
    \State Update actor by maximizing the clipped PPO objective (Equation~\eqref{eq:ppo_loss}), where $\pi^{\text{old}}$ uses stored $\log\pi_k$ from $\mathcal{T}$
    \State Update critic by minimizing the MSE loss (Eq.~\eqref{eq:ppo_critic})
\EndFor
\end{algorithmic}
\end{algorithm}

\noindent\textit{Intuition.}  
The PPO framework enables the agent to learn dose-aware acquisition strategies directly through interaction with the CT environment. 
Rather than distributing photons uniformly across all projection angles,
the learned policy adaptively selects which angles to measure at each step.
Since a fixed dose \(\Delta b\) is used for every acquisition, repeatedly selecting
the same angle results in a higher cumulative dose being allocated to that
direction, while angles that are selected fewer times receive less total dose.
In this way, the policy implicitly concentrates the dose budget on more
informative views through repeated measurements, without reallocating or
reducing the dose that has already been acquired.
The critic provides value estimates that stabilize training, and the reward function drives the policy to optimize either overall reconstruction quality (PSNR) or defect detectability (CNR). 
This design balances exploration of new angles with exploitation of informative ones, leading to improved reconstruction quality under limited-dose conditions.

\begin{figure}[H]
  \centering
  \includegraphics[width=\linewidth]{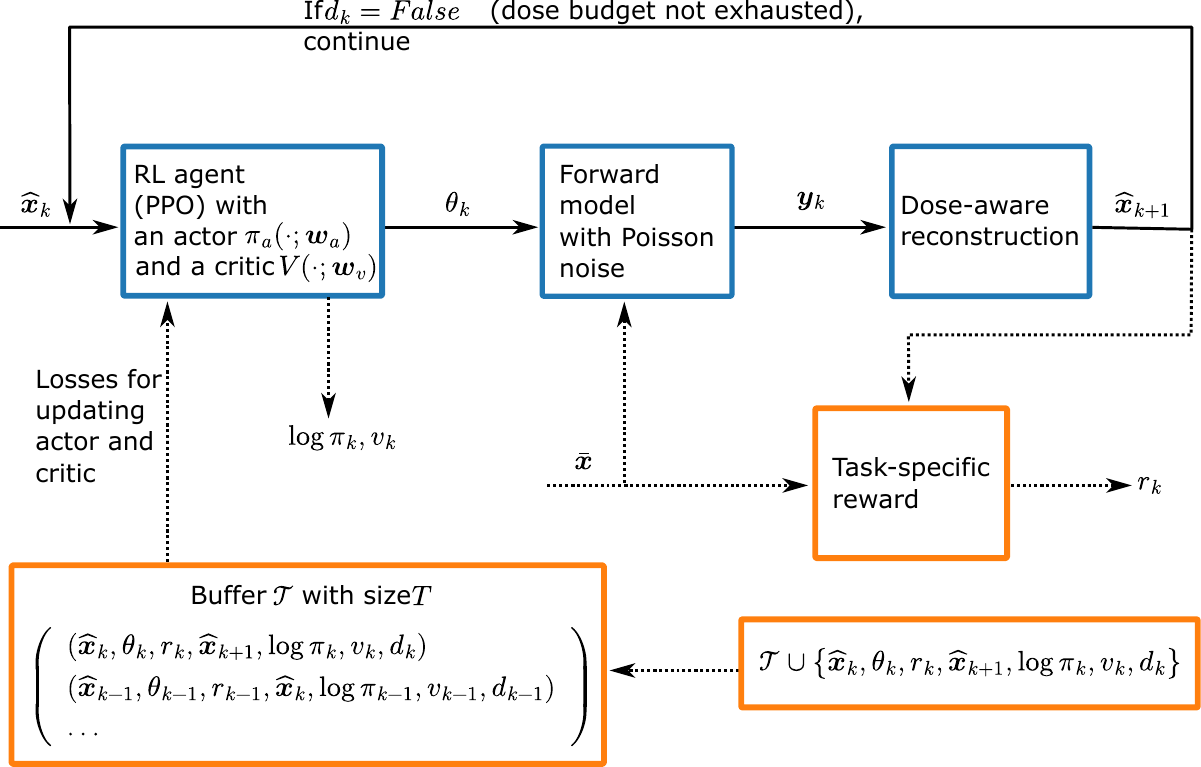}
  \caption{Pipeline of the proposed method. 
  At each step $k$, the dose-aware reconstruction produces the current state $\widehat{\boldsymbol{x}}_{k}$. 
  The PPO agent selects an angle $\theta_k$ based on the current policy $\pi_{k}$, and the environment returns the next state $\widehat{\boldsymbol{x}}_{k+1}$ and reward $r_k$. 
  The transition $(\widehat{\boldsymbol{x}}_{k}, \theta_k, r_k, \widehat{\boldsymbol{x}}_{k+1}, \log \pi_k, v_k, d_k)$ is stored in the buffer $\mathcal{T}$ for PPO updates. 
  The pipeline continues until the dose budget is exhausted. 
  Orange blocks and dashed arrows denote components used only during training, whereas in real applications only the blue blocks and solid arrows are executed.}
  \label{fig:pipeline}
\end{figure}

The overall pipeline is illustrated in Figure~(\ref{fig:pipeline}). 
At each step $k$, the dose-aware reconstruction produces the current state $\widehat{\boldsymbol{x}}_{k}$. 
The PPO agent then selects an angle $\theta_k$ based on the current policy $\pi_{k}$, and the environment returns the next state $\widehat{\boldsymbol{x}}_{k+1}$ along with the reward $r_k$. 
The transition $(\widehat{\boldsymbol{x}}_{k}, \theta_k, r_k, \widehat{\boldsymbol{x}}_{k+1}, \log \pi_k, v_k, d_k)$ is appended to the buffer $\mathcal{T}$ for PPO updates. 
The pipeline continues until the dose budget is exhausted. 
Orange blocks and dashed arrows denote components used only during training, whereas in real applications only the blue blocks and solid arrows are executed.

\section{Results}
\label{sec:results}
We present experimental results to validate both components of the proposed framework. 
First, we evaluate the dose-aware PnP–PWLS reconstruction algorithm against standard iterative methods. 
Next, we assess the RL policy for joint angle selection and dose allocation, using the golden ratio policy as a baseline. 
Finally, we demonstrate the effectiveness of the proposed approach on a specific task of interest—defect detection.





\subsection{Datasets}
\label{sec:dataset}

\begin{figure}[H]
  \centering
  \includegraphics[width=\linewidth]{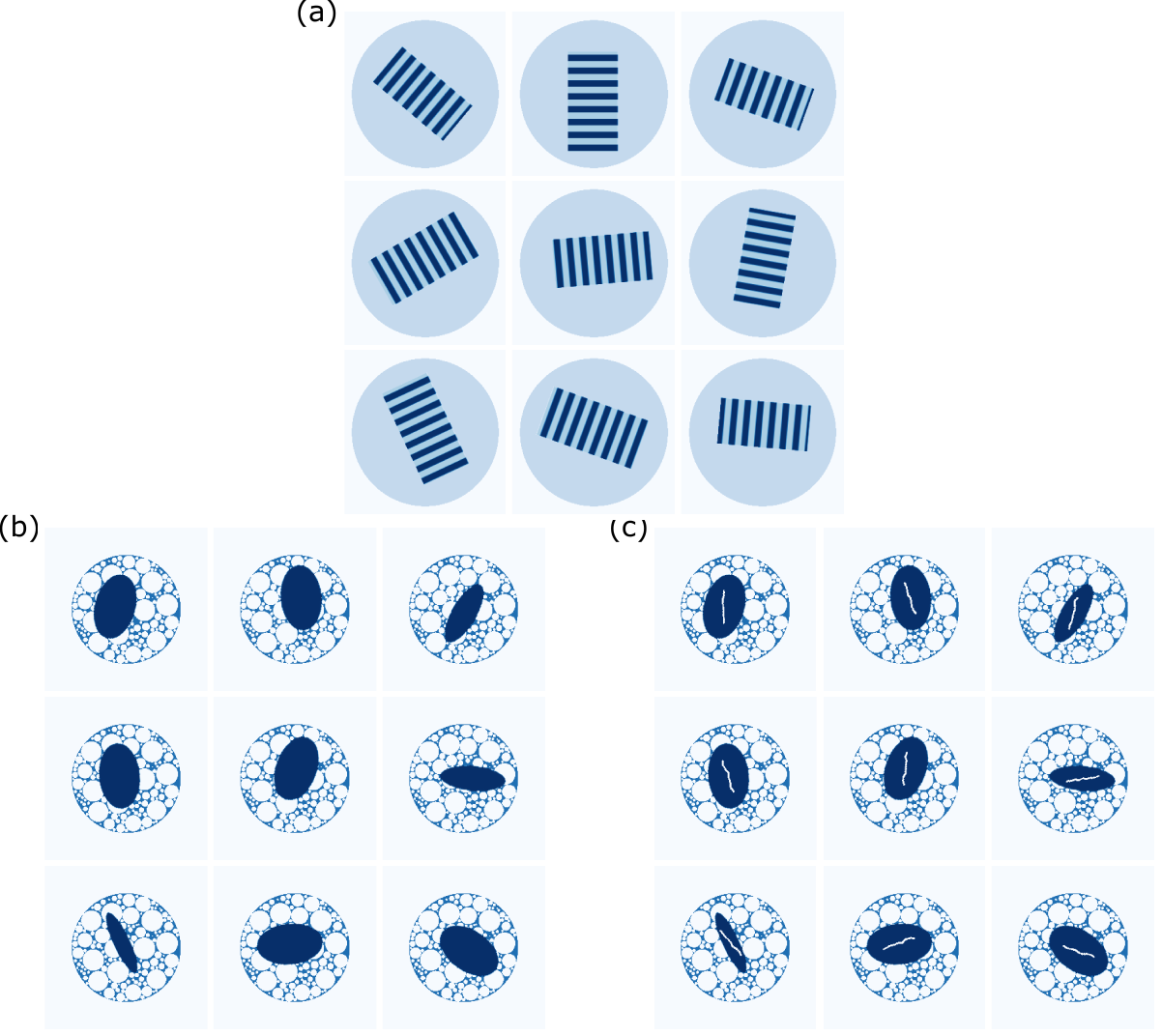}
    \caption{Example training dataset consisting of three types of phantoms. 
    (a) Wedge-shaped phantoms composed of two different materials, with variations in rotation, position, and size. 
    (b) Ellipse- and foam-based phantoms with different material properties. 
    (c) Ellipse-based phantoms containing defects. 
    The ellipses vary in rotation, position, size, and roundness, providing a diverse set of training samples.}

  \label{fig:Trainingdata}
\end{figure}

In our numerical experiments, we consider several phantoms illustrated in Figure~(\ref{fig:Trainingdata}), each with a size of $256 \times 256$ pixels. 
The phantoms are composed of multiple materials and grouped into three datasets. 
Dataset (a) contains wedge-shaped phantoms composed of different materials, with variations in rotation, position, and size. These phantoms are inspired by
industrial CT inspection scenarios involving fiber-reinforced or layered composite
materials, where internal structures exhibit strong directional characteristics.
Datasets (b) and (c) consist of ellipse- and foam-based phantoms \cite{pelt2022foam} with different material properties. These phantoms are inspired by polymer or metal foam–like materials that
act as containers for embedded foreign objects or defects with preferential
orientations. Such configurations are representative of industrial CT
inspection tasks, where the goal is to detect and characterize internal
objects or anisotropic defects within a heterogeneous, porous background.
In particular, dataset (c) includes ellipse-based phantoms containing defects. 
All datasets include phantoms at different rotations, represented by 36 equally spaced angles between $0^{\circ}$ and $179^{\circ}$. 
In addition, the phantoms exhibit variations in scaling and spatial shifts.

\subsection{Implementation}
Tomosipo \cite{hendriksen2021tomosipo} was used to generate simulated projections and perform reconstructions. 
The image size was set to $256 \times 256$, and a projection size of $384$ was chosen to ensure that each view fully covered the image domain.  

For the reconstruction algorithm, the parameter \(h\) in Algorithm~\ref{alg:dose_weighted_PnP_reconstruction} was set to \(1.8\), and the PnP--TV denoiser employed a fixed proximal step size of \(0.0005\). 
In the RL framework (Algorithm~\ref{alg:ppo-ct}), the discount factor was set to \(\gamma = 0.99\), and the replay buffer capacity was fixed at \(500\) transitions. 
The actor and critic losses were weighted by \(1.0\) and \(0.5\), respectively, and an entropy regularization term with weight \(0.01\) was included to promote adequate exploration during training (see Appendix~\ref{appendix:ppo} for details). 
These hyperparameters were selected empirically to provide a stable balance between policy optimization, accurate value-function estimation, and robust exploratory behavior. 
A convolutional encoder was employed to extract low-dimensional feature representations from the input images, following the design in~\cite{wang2024sequential}. 
All network parameters were optimized using the Adam algorithm~\cite{kingma2014adam} with a learning rate of \(10^{-4}\) and a weight decay of \(10^{-5}\).

\subsection{Experiment 1: Reconstruction Using Dose-Aware PnP–PWLS with a TV Prior}

This experiment evaluates the proposed reconstruction algorithm against standard iterative methods.  
We focus on datasets (a) and (b) described in Section~\ref{sec:dataset}. Reconstructions were performed with three equal angular sampling settings: 36, 48, and 60 angles. To simulate unequal dose allocation across views, odd-indexed angles were assigned a low dose (100 photons), and even-indexed angles a high dose (1000 photons). The number of gradient descent iterations was varied from 5 to 100 in increments of 5.  

Four methods were compared: Simultaneous Iterative Reconstruction Technique (SIRT), conventional PWLS, dose-aware PWLS, and dose-aware PnP–PWLS with a TV prior. All reconstructions were performed under a non-negativity constraint.  

\begin{figure}[H]
  \centering
  \includegraphics[width=\linewidth]{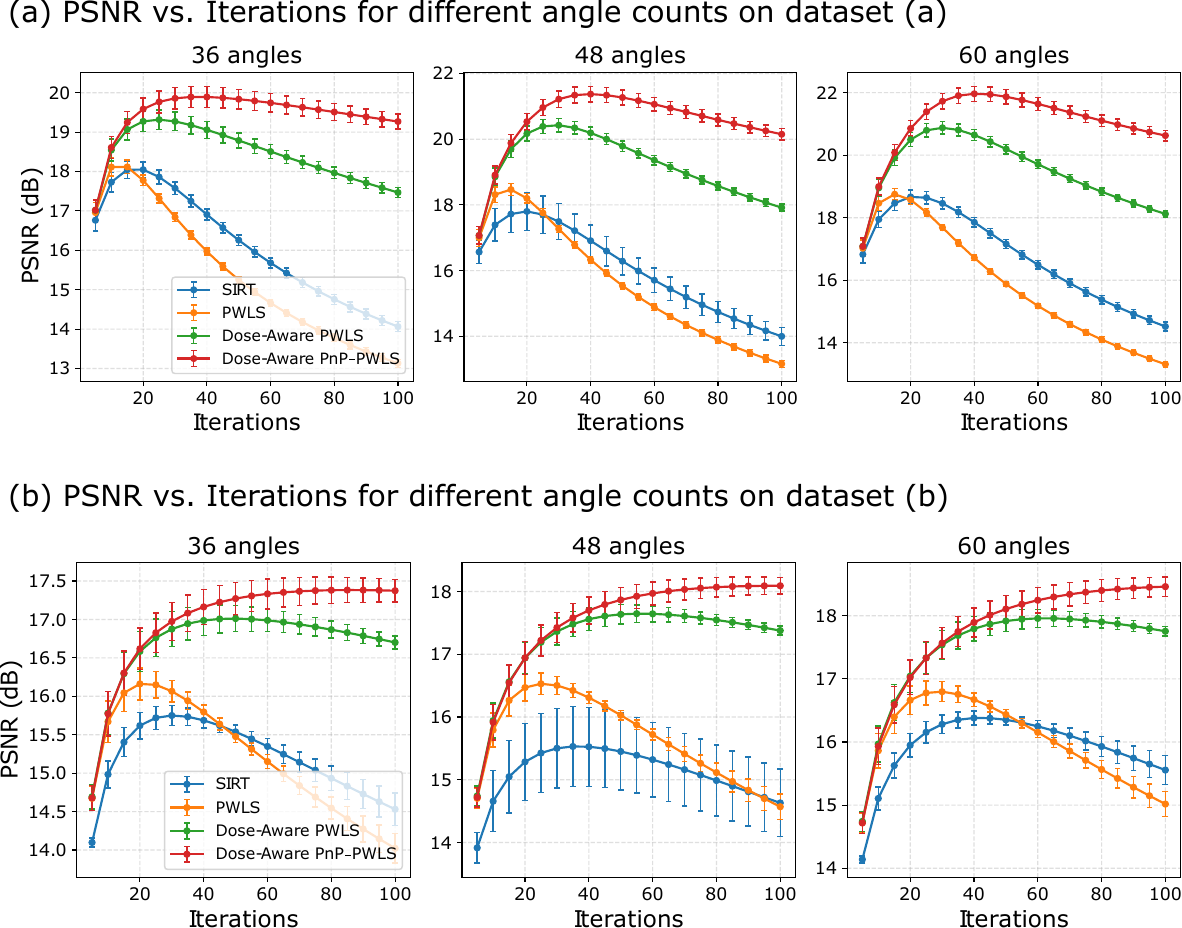}
  \caption{PSNR versus number of iterations for different angular sampling settings (36, 48, and 60 angles, left to right) on datasets (a) and (b). The four methods compared are SIRT, conventional PWLS, dose-aware PWLS, and dose-aware PnP–PWLS. Curves show the mean PSNR across all non-defective phantoms in the dataset, and error bars indicate the standard deviation.}
  \label{fig:PSNR_comparison}
\end{figure}

Figure~(\ref{fig:PSNR_comparison}) compares reconstruction performance on datasets (a) and (b). Both dose-aware variants (PWLS and PnP–PWLS with a TV prior) consistently achieve higher PSNR and exhibit greater robustness to the number of iterations across different angular sampling settings. In contrast, SIRT and conventional PWLS degrade noticeably at higher iteration counts, making reconstruction quality more sensitive to the iteration choice and less reliable for subsequent experiments. These results demonstrate that dose-aware weighting improves reconstruction stability, with PnP–PWLS with a TV prior consistently providing the best overall performance.  

\noindent\textit{Remark.}  
The performance gap between the proposed dose-aware methods and the baselines becomes more pronounced at higher iteration counts. While SIRT and conventional PWLS tend to overfit the noise and degrade in quality, dose-aware weighting preserves relative stability by scaling the contribution of each view according to its photon statistics, although some degradation remains visible on dataset (a). This highlights that dose-awareness is particularly beneficial in long-iteration regimes, where conventional methods become unreliable. In addition, incorporating a TV prior provides effective denoising while preserving edges, further improving reconstruction quality.

\begin{figure}[H]
  \centering
  \includegraphics[width=\linewidth]{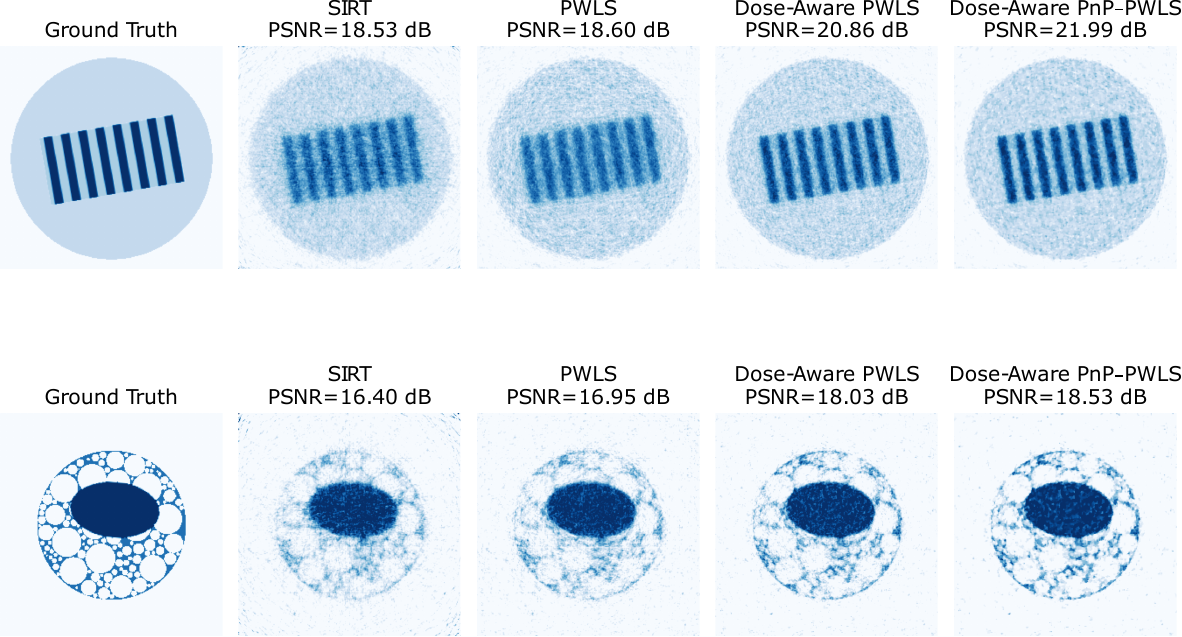}
  \caption{Reconstruction results for two phantoms with 60 projection angles using SIRT, conventional PWLS, and the proposed dose-aware methods. For each method, the iteration count was selected from 0 to 100 to yield its best reconstruction. SIRT reconstructions suffer from low contrast and artifacts, while PWLS provides modest improvements but remains noisy. The proposed dose-aware approaches achieve the highest PSNR and visibly improve contrast.}
  \label{fig:Rec_sample}
\end{figure}

These visual results are consistent with the quantitative PSNR values reported in Figure~(\ref{fig:Rec_sample}), where the proposed dose-aware methods achieve improvements of approximately 1.5–3 dB over SIRT and conventional PWLS when considering their best reconstructions within the iteration range of 0–100. This demonstrates that explicitly modeling angle-dependent dose allocation not only enhances numerical performance but also leads to visibly cleaner and higher-contrast reconstructions.

\subsubsection{Experiment 2: Angle Selection and Dose Allocation for Reconstruction Quality Improvement}

This experiment evaluates whether the RL agent can jointly select projection angles and allocate dose to improve reconstruction quality when the reward function is based solely on PSNR. 
We used dataset (a) for this study and compared the learned policy against the \emph{Golden Ratio (GR) policy} \cite{kohler2004projection,craig2023real}, which is a common baseline in sparse-view CT. 
The GR policy generates angles by applying an irrational angular increment, resulting in a non-repeating sequence that uniformly fills the angular space over time. Following the previous experiments, we adopt the most effective reconstruction algorithm, namely the dose-aware PnP–PWLS with a TV prior.

\paragraph{Training procedure.} 
The agent was trained on 150 phantoms from dataset (a). 
For the GR policy, 20 angles were used with an equal dose assigned to each angle. 
Three separate RL policies
were trained for three different total photon budgets of 6000 photons (300 photons per angle), 7000 photons (350 photons per angle) and 8000 photons (400 photons per angle). 
For the RL policy, the agent could allocate more dose to informative angles by repeating them. 
For example, with a base dose of 300 photons per angle, if angle $a$ is repeated $n$ times, the total dose assigned to that angle becomes $n \times 300$. 
Training was performed for 1500 episodes using the reward function in Eq.~\eqref{eq:rewardfunction}, assuming that the ground truth $\bar{\boldsymbol{x}}$ was non-defective. 
The weight parameter $\alpha$ in Eq.~\eqref{eq:rewardfunction} was set to 0.2, and the RoI $\Omega$ corresponded to the full image.

\paragraph{Validation procedure.} 
Validation was performed on a separate test set of 100 phantoms with unseen rotation angles. 
All results reported in this section are obtained from this validation set using the trained policy.

\paragraph{Quantitative comparison.}
Table~(\ref{tb:comparison_wedge}) summarizes the performance of the GR and RL policies on the unseen-rotation test set across these dose levels (6000, 7000, and 8000 photons).
The RL policy achieves a higher PSNR value than the GR policy, thereby preserving overall reconstruction quality. As the increment of the total photons, the number of angles and PSNR values also increase, meaning that the repetition of angle is reduced due to more budget.

\begin{table}[H]
\caption{Performance comparison of GR policy and RL policy on unseen rotations tested at different dose levels on dataset (a). Results are reported as mean $\pm$ standard deviation.}

\centering
\begin{tabular}{c|c|c|c}
\hline
\textbf{Policies} & \textbf{Total photons} & \textbf{PSNR} & \makecell{\textbf{Average} \\ \textbf{number of angles}} \\ 
\hline
\multirow{3}{*}{GR policy} 
 & 6000 & 16.82 $\pm$ 0.39 & 20 \\
 & 7000 & 17.26 $\pm$ 0.44 & 20 \\
 & 8000 & 17.61 $\pm$ 0.48 & 20 \\

\hline
\multirow{3}{*}{RL policy} 
 & 6000 & 19.52 $\pm$ 0.76 & 7.32 $\pm$ 1.04\\
 & 7000 & 19.83 $\pm$ 0.74 & 8.05 $\pm$ 1.50 \\
 & 8000 & 20.33 $\pm$ 0.58 & 8.82 $\pm$ 1.20 \\
 
\hline
\end{tabular}
\label{tb:comparison_wedge}
\end{table}

\begin{figure}[H]
\centering
\includegraphics[width=\linewidth]{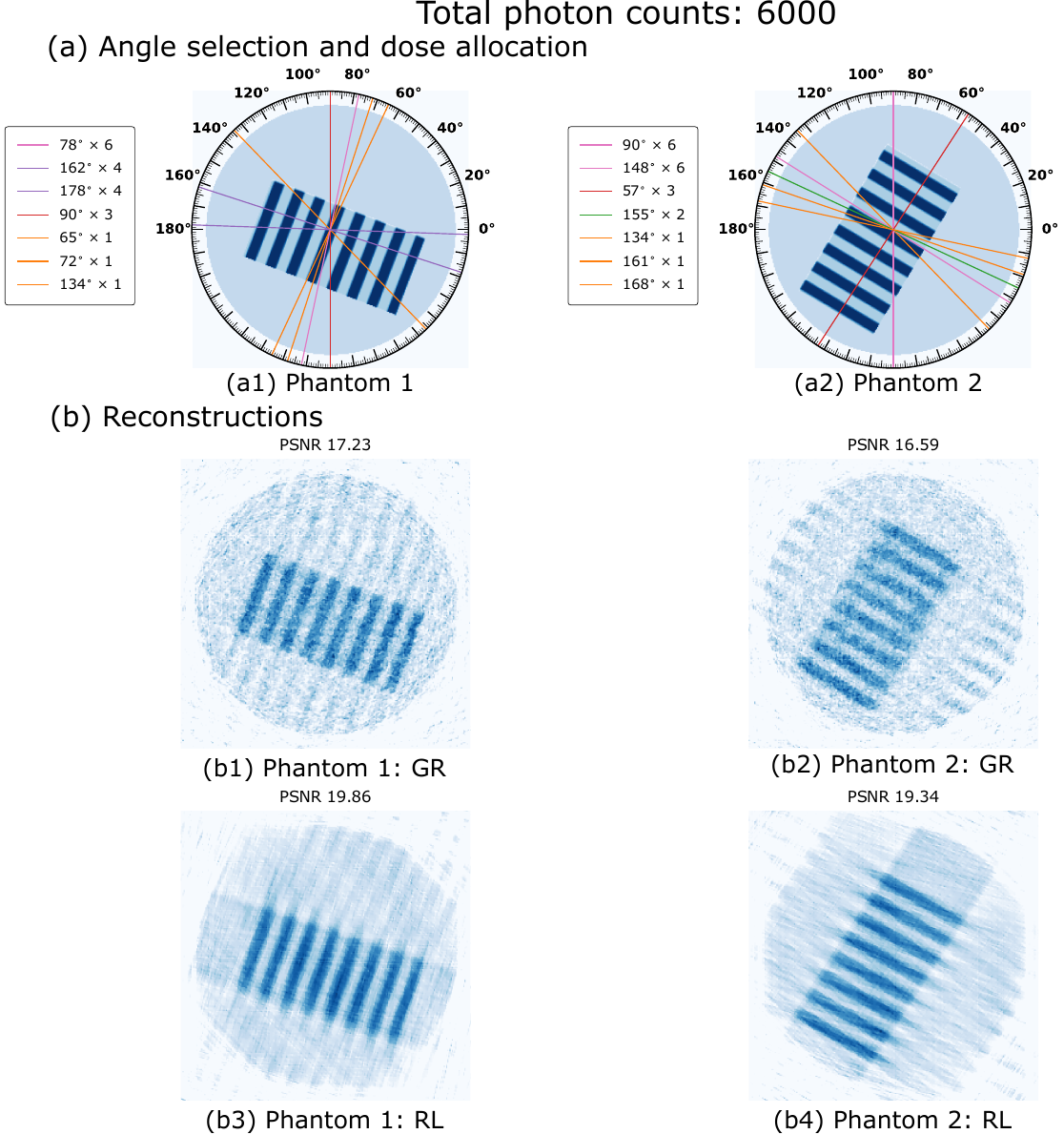}
\caption{
(a) Angle selection and dose allocation at a total dose of 6000 photons for wedge-shaped phantoms (a1–a2). 
Each colored spoke represents a parallel-beam view at angle~$\theta$, measured counter-clockwise from the $x$-axis; angles $\theta$ and $\theta+180^\circ$ are equivalent. 
Colors indicate the number of repeated selections and thus the allocated photons at that angle. 
With 6000 photons in total, each selection corresponds to 300 photons; for example, an orange spoke denotes one selection (300 photons), and a green spoke denotes two selections (600 photons). 
(b) Reconstructions obtained under the GR (b1–b2) and RL (b3–b4) policies at the same total dose. 
The RL policy adaptively concentrates measurements in more informative directions, yielding improved reconstruction quality.
}
\label{fig:angle_selection_wedge6000}
\end{figure}

\begin{figure}[H]
\centering
\includegraphics[width=\linewidth]{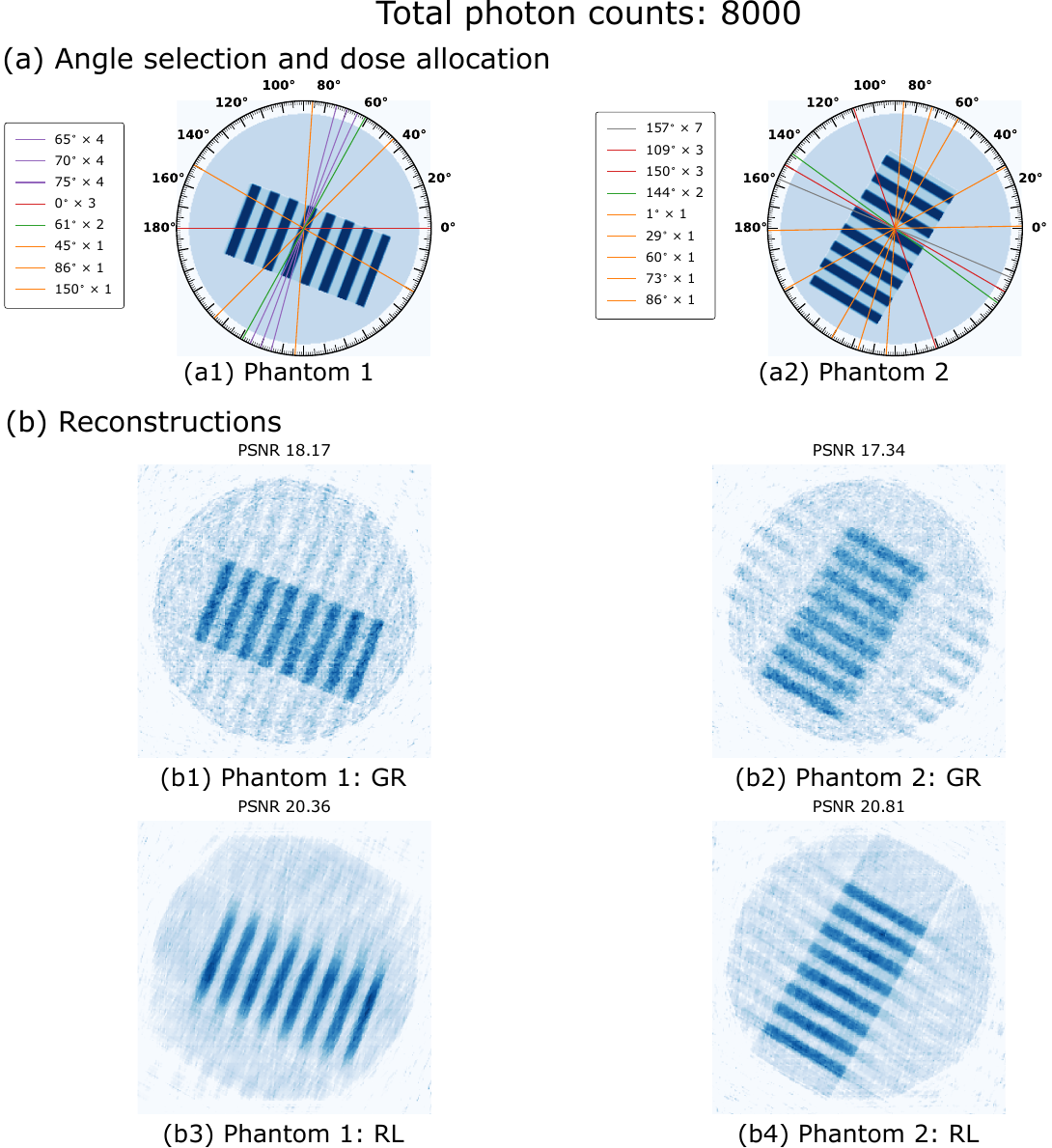}
\caption{
(a) Angle selection and dose allocation at a total dose of 8000 photons for wedge-shaped phantoms (a1–a2). 
Each colored spoke represents a parallel-beam view at angle~$\theta$, measured counter-clockwise from the $x$-axis; angles $\theta$ and $\theta+180^\circ$ are equivalent. 
Colors indicate the number of repeated selections and thus the allocated photons at that angle. 
With 8000 photons in total, each selection corresponds to 400 photons; for example, an orange spoke denotes one selection (400 photons), and a green spoke denotes two selections (800 photons). 
(b) Reconstructions obtained under the GR (b1–b2) and RL (b3–b4) policies at the same total dose. 
The RL policy adaptively concentrates measurements in more informative directions, yielding improved reconstruction quality.
}
\label{fig:angle_selection_wedge8000}
\end{figure}

\paragraph{Angle selection and reconstruction analysis.}  
Figures~(\ref{fig:angle_selection_wedge6000}) and~(\ref{fig:angle_selection_wedge8000}) (a1–a2) illustrate the angle selections produced by the GR and RL policies for wedge-shaped phantoms at total doses of 6000 and 8000 photons, respectively. 
Each colored spoke denotes the beam direction of a parallel-beam view at angle $\theta$ (degrees), measured counter-clockwise from the $x$-axis, with $\theta$ and $\theta+180^\circ$ being equivalent. 
The color coding indicates the number of repeats, corresponding to the relative photon allocation at that angle.
The GR policy distributes views uniformly across the angular range, whereas the RL policy adapts its strategy by concentrating angles and dose around preferential directions corresponding to the wedge orientations. 
This adaptive allocation provides more informative measurements along high-contrast directions while reducing sampling in less informative regions.

Figures~(\ref{fig:angle_selection_wedge6000}) and~(\ref{fig:angle_selection_wedge8000}) (b1–b4) present reconstruction results for wedge-shaped phantoms using the GR and RL policies at total doses of 6000 and 8000 photons. 
For both dose levels, reconstructions obtained with the GR policy appear noisy and exhibit poor contrast between materials. 
In contrast, the RL policy produces reconstructions with clearer separation of material regions and reduced noise, demonstrating the advantage of adaptive angle selection and dose allocation.

\subsubsection{Experiment 3: Angle Selection and Dose Allocation for Defect Detectability Improvement}
\begin{figure}[H]
  \centering
  \includegraphics[width=\linewidth]{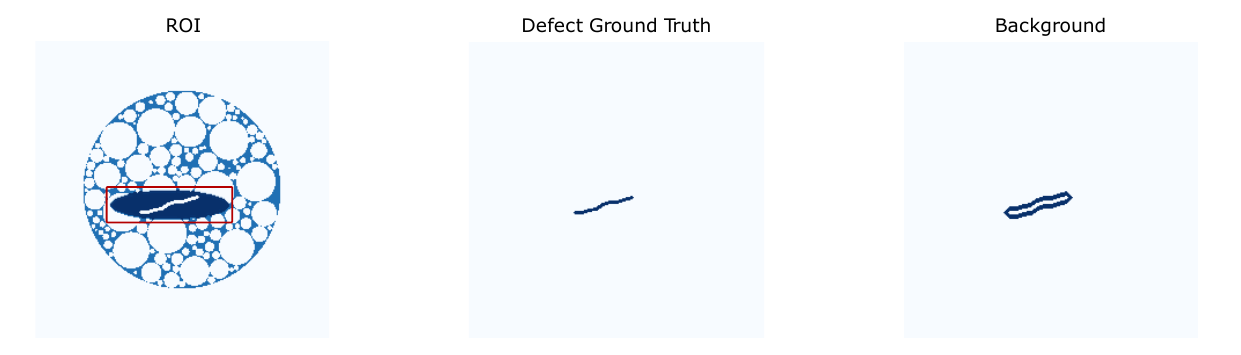}
  \caption{Illustration of the regions used for defect detection. (Left) ROI highlighted by a red rectangle. (Middle) Ground-truth defect mask. (Right) Background region used for CNR calculation.}
  \label{fig:ground_truth_defect}
\end{figure}

This experiment addresses the task of defect detection. Both dataset (b) (non-defective) and dataset (c) (defective) phantoms were used to evaluate whether the agent can select angles and allocate dose in a way that maintains reconstruction quality for non-defective phantoms while improving defect detectability. We still used the dose-aware PnP–PWLS with a TV prior as the reconstruction algorithm. The elliptical region is assumed to be defect-prone, and the ground-truth defect mask is used during training. Accordingly, we define an ROI covering the ellipse and compute rewards within this region. Figure (\ref{fig:ground_truth_defect}) illustrates the ROI, the ground-truth defect mask, and the surrounding background region used for computing the CNR in Equation~\eqref{eq:CNR}.

\paragraph{Training procedure.} 
The agent was trained on a dataset of 300 phantoms, comprising 150 non-defective and 150 defective cases. 
Three separate RL policies were trained for three different total photon budgets of 1000, 1500, and 2000.  
Training was performed for 2000 episodes using the reward function defined in Eq.~\eqref{eq:rewardfunction}. The weight parameters $\alpha$ and $\beta$ in Eq.~\eqref{eq:rewardfunction} were set to 0.2 and 3.5, respectively, to balance the two tasks.
\begin{equation}
    R(\bar{\boldsymbol{x}}, \widehat{\boldsymbol{x}}) =
    \begin{cases}
      \alpha \,\mathrm{PSNR}_{\Omega}(\bar{\boldsymbol{x}}, \widehat{\boldsymbol{x}}), & \text{if $\bar{\boldsymbol{x}}$ is non-defective}, \\[6pt]
      \beta \,\mathrm{CNR}_{\mathcal{D}, \mathcal{B}}(\bar{\boldsymbol{x}}, \widehat{\boldsymbol{x}}), & \text{if $\bar{\boldsymbol{x}}$ is defective},
    \end{cases}
\label{eq:rewardfunction}
\end{equation}
where $\alpha$ and $\beta$ control the relative importance of reconstruction accuracy and defect detectability.

\paragraph{Validation procedure.} 
Evaluation was carried out on a separate test set of 200 phantoms with unseen rotation angles, including 100 non-defective and 100 defective cases. 
All results reported in this section are obtained from this validation set using the trained policy.  

\paragraph{Quantitative comparison.}  
Table~(\ref{tb:comparison_defect}) summarizes the performance of the GR and RL policies on the unseen-rotation test set across three dose levels (1000, 1500, and 2000 photons). 
For non-defective phantoms, the RL policy achieves PSNR comparable to the GR policy, thereby preserving overall reconstruction quality. 
For defective phantoms, the RL policy consistently outperforms the GR policy in terms of CNR at all dose levels, demonstrating its ability to allocate dose more effectively for defect detection. 
Consistent with the previous experiment, increasing the total photon budget results in a larger number of distinct angles being selected, thereby reducing the need for repeated angles.

\begin{table}[H]
\caption{Performance comparison of GR and RL policies on unseen-rotation tested at different dose levels on datasets (a) and (b). Results are mean $\pm$ standard deviation.}
\centering
\begin{tabular}{c|c|c|c|c}
\hline
\textbf{Policy} & \textbf{Total photons} & \makecell{\textbf{PSNR}\\(Non-def.)} & \makecell{\textbf{CNR}\\(Def.)} & \makecell{\textbf{Average} \\ \textbf{number of angles}} \\ 
\hline
\multirow{3}{*}{GR policy} 
 & 1000 & 19.51 $\pm$ 1.29 & 0.51 $\pm$ 0.26 & 10 \\
 & 1500 & 20.68 $\pm$ 1.21 & 0.61 $\pm$ 0.27 & 10 \\
 & 2000 & 21.38 $\pm$ 1.19 & 0.68 $\pm$ 0.29 & 10 \\
\hline
\multirow{3}{*}{RL policy} 
 & 1000 & 20.04 $\pm$ 1.48 & 1.01 $\pm$ 0.42 & 6.90 $\pm$ 1.40 \\
 & 1500 & 20.77 $\pm$ 1.47 & 1.31 $\pm$ 0.41 & 7.13 $\pm$ 1.18 \\
 & 2000 & 21.06 $\pm$ 1.46 & 1.33 $\pm$ 0.43 & 7.54 $\pm$ 1.25 \\
\hline
\end{tabular}
\label{tb:comparison_defect}
\end{table}

\begin{figure}[H]
\centering
\includegraphics[width=\linewidth]{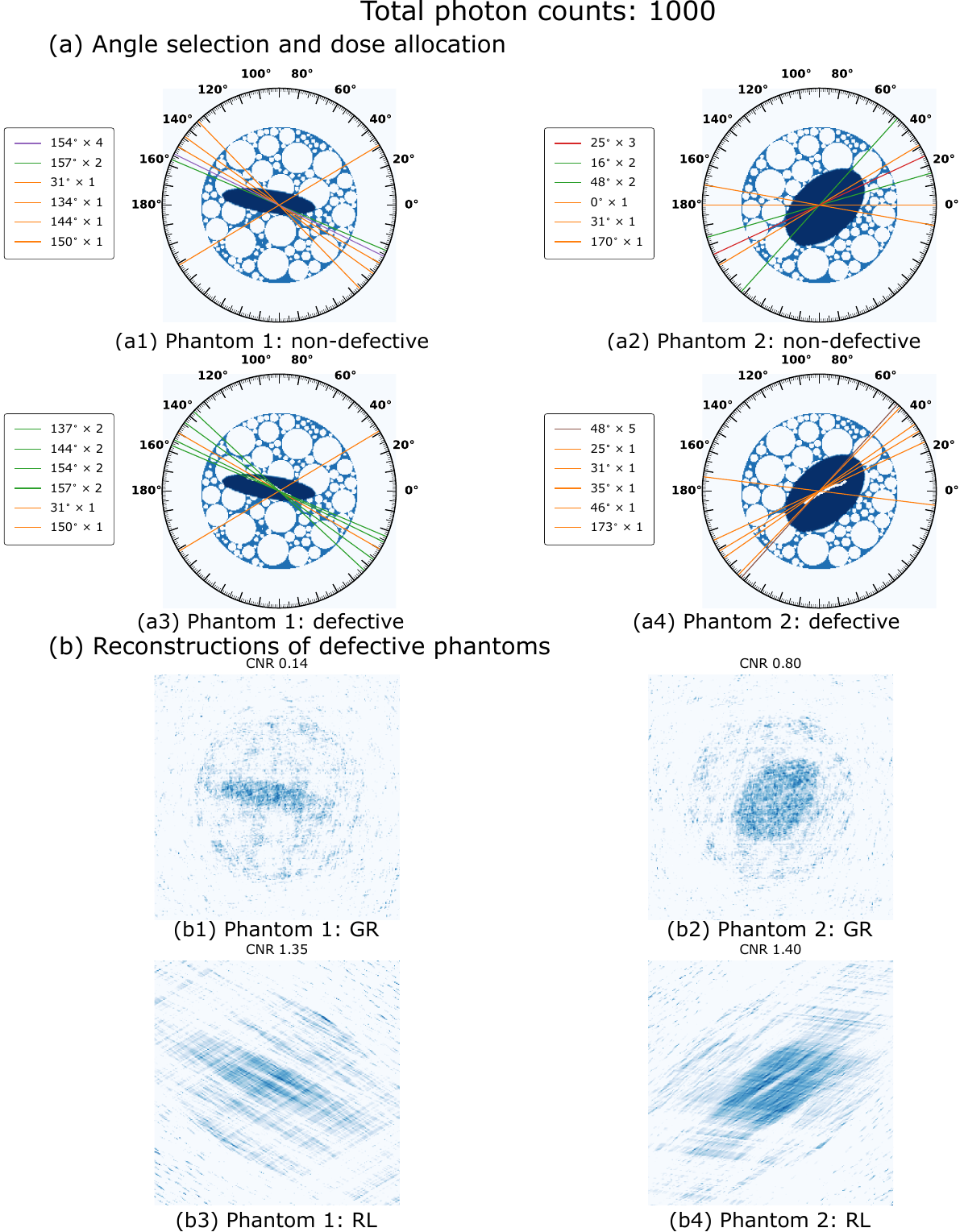}
\caption{
(a) Angle selection and dose allocation at a total dose of 1000 photons for non-defective (a1–a2) and defective (a3–a4) phantoms. 
Colored spokes denote projection angles, with color indicating the number of repeats (i.e., relative photon allocation). 
An orange spoke corresponds to one selection (base level: 100 photons), while colors such as green indicate multiple repeats and therefore higher allocated photons.  
(b) Reconstructions of defective phantoms using the GR (b1–b2) and RL (b3–b4) policies. 
The RL policy adaptively concentrates measurements near the defect, yielding higher CNR and improved defect visibility compared with the fixed GR policy.
}

\label{fig:angle_selection_1000}
\end{figure}

\begin{figure}[H]
\centering
\includegraphics[width=\linewidth]{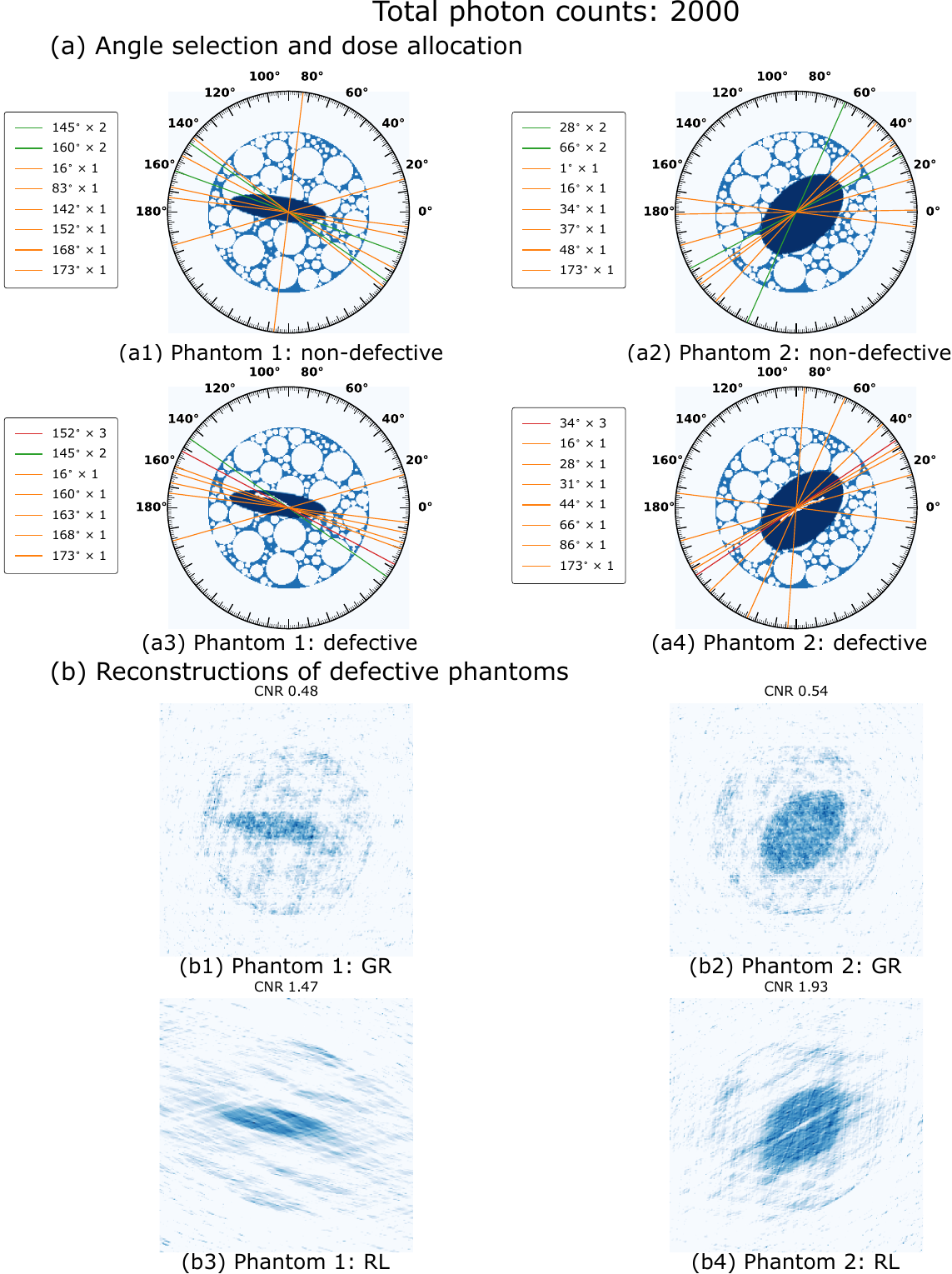}
\caption{
(a) Angle selection and dose allocation at a total dose of 2000 photons for non-defective (a1–a2) and defective (a3–a4) phantoms. 
Colored spokes denote projection angles, with color indicating the number of repeats (i.e., relative photon allocation). 
An orange spoke corresponds to one selection (base level: 200 photons), while colors such as green indicate multiple repeats and therefore higher allocated photons.  
(b) Reconstructions of defective phantoms using the GR (b1–b2) and RL (b3–b4) policies. 
The RL policy adaptively concentrates measurements near the defect, yielding higher CNR and improved defect visibility compared with the fixed GR policy.
}
\label{fig:angle_selection_2000}
\end{figure}

\paragraph{Angle selection and reconstruction analysis.}
Figures~(\ref{fig:angle_selection_1000}) and~(\ref{fig:angle_selection_2000}) (a1-a4) illustrate the angle selections produced by the GR and RL policies for non-defective and defective phantoms at total doses of 1000 and 2000 photons, respectively.
While the GR policy distributes views uniformly regardless of phantom type, the RL policy adapts its strategy: for defective phantoms, angles and dose are concentrated around the defect region.
Comparing across dose levels shows that higher photon budgets lead to broader angular coverage, while the presence of defects drives the RL policy to focus sampling more locally.

The ultimate goal of this experiment is to evaluate whether adaptive angle and dose selection can improve defect detectability.
Figures~(\ref{fig:angle_selection_1000}) and~(\ref{fig:angle_selection_2000}) (b1-b4) show reconstruction results for defective phantoms at total doses of 1000 and 2000 photons, respectively.
Reconstructions using the GR policy fail to make the defects clearly visible.
In contrast, reconstructions obtained with the RL policy consistently achieve higher CNR values and reveal defects more distinctly, validating the benefit of dose-aware adaptive view selection for defect detection.

\newpage

\noindent\textbf{Summary.}
These experiments demonstrate that the proposed RL policy adaptively selects relative informative angles and allocates dose across projection angles according to phantom content.
It preserves reconstruction quality for non-defective phantoms while substantially improving defect detectability in defective cases, consistently outperforming the GR policy in both CNR and visual clarity.

\section{Discussion}
\label{sec:discussion}
These results demonstrate that the proposed dose-aware reconstruction algorithms are particularly effective when dose is distributed unequally across projection angles. 
Moreover, the RL policy achieves superior reconstruction quality and defect detectability by leveraging fewer angles and adaptively distributing dose within a fixed budget, compared to the GR policy, which relies on more angles with equal dose allocation.  

Despite these encouraging findings, several limitations remain. 
First, only a Poisson noise model was considered, whereas real CT measurements are often better described by a mixture of Gaussian and Poisson noise \cite{andriiashen2024x, graas2025scintillator}. 
Accurate modeling of the noise characteristics of a specific CT system will be important for translating this approach to real data. 
Second, extending the framework to three-dimensional datasets introduces additional challenges: iterative reconstruction becomes more computationally demanding, and the state and action spaces for RL grow substantially. 
Finally, future work may explore the integration of downstream tasks such as segmentation and classification, which could further highlight the practical benefits of adaptive angle and dose allocation.

\section{Conclusion}
\label{sec:conclusion}
In this paper, we proposed dose-aware reconstruction methods and developed a reinforcement learning strategy for joint angle selection and dose allocation. 
Our results demonstrate that explicitly modeling angle-dependent dose variation improves reconstruction quality and defect detectability under unequal dose distributions. 
Furthermore, the proposed RL-based framework shows the potential to reduce the number of projection angles while maintaining or improving image quality, thereby enabling more efficient and task-adaptive CT scanning.

\section*{Acknowledgement}

The authors gratefully acknowledge the support of the xCTing project. We also acknowledge the use of OpenAI’s large language model, ChatGPT based on GPT-4o, to assist in refining the manuscript text. The tool was used at the sentence level for tasks such as grammar correction and sentence rephrasing.

\section*{Author contributions}

\textbf{Tianyuan Wang}: Methodology, Visualization, Software, Writing -- original draft; Writing -- review \& editing; \textbf{Dani\"el M. Pelt}: Conceptualization, Visualization, Writing -- review \& editing; \textbf{Felix Lucka}: Investigation, Validation, Writing -- review \& editing; \textbf{Tristan van Leeuwen}: Conceptualization, Supervision, Validation; \textbf{K. Joost Batenburg}: Conceptualization, Project administration, Resources.

\section*{Disclosure statement}
No potential conflict of interest was reported by the author(s).

\section*{Funding}
This research was co-financed by the European Union H2020-MSCA-ITN-2020 under grant agreement no. 956172 (xCTing).

\section*{Data availability statement}
The implementation and synthetic datasets used in this study will be made publicly available on GitHub: https://github.com/tianyuan1wang/ct-angle-dose-rl.

\newpage
\appendix

\section{Power Iteration for Lipschitz Constant Estimation}
\label{sec:power-iteration}

To ensure stable gradient descent updates in Algorithm~\ref{alg:dose_weighted_PnP_reconstruction}, 
the step size is normalized by the Lipschitz constant 
\[
L = \|\boldsymbol{A}^\top \boldsymbol{W} \boldsymbol{A}\| ,
\] 
which corresponds to the spectral norm of the weighted normal operator.  
Since exact computation of \(L\) is infeasible for large-scale CT systems, we approximate it using the standard \emph{power iteration} method.  
The procedure is summarized in Algorithm~\ref{alg:power-iteration}.

\begin{algorithm}[H]
\caption{Power Iteration to Estimate Spectral Norm \(L = \|\boldsymbol{A}^\top \boldsymbol{W} \boldsymbol{A}\|\)}
\label{alg:power-iteration}
\begin{algorithmic}[1]
\Require 
Forward operator $\boldsymbol{A}(\boldsymbol{\theta})$;  
transpose operator $\boldsymbol{A}^\top(\boldsymbol{\theta})$;  
weighting operator $\mathcal{W}(\cdot)$;  
domain size $n$;  
maximum iterations $T$.
\Ensure 
Spectral norm estimate $L$.

\vspace{0.25em}
\State Initialize: $\boldsymbol{z}^{(0)} \sim \mathcal{N}(\boldsymbol{0}, \mathbf{I}_n)$; normalize $\boldsymbol{z}^{(0)} \gets \boldsymbol{z}^{(0)} / \|\boldsymbol{z}^{(0)}\|$
\For{$t = 0$ \textbf{to} $T-1$}
    \State $\boldsymbol{u} \gets \boldsymbol{A}(\boldsymbol{\theta}) \, \boldsymbol{z}^{(t)}$ \hfill \Comment{Forward projection}
    \State $\boldsymbol{v} \gets \mathcal{W}(\boldsymbol{u})$ \hfill \Comment{Dose-aware weighting}
    \State $\boldsymbol{w} \gets \boldsymbol{A}^\top(\boldsymbol{\theta}) \, \boldsymbol{v}$ \hfill \Comment{Backprojection}
    \State Normalize: $\boldsymbol{z}^{(t+1)} \gets \boldsymbol{w} / \|\boldsymbol{w}\|$
    \State Update estimate: $L^{(t+1)} \gets \|\boldsymbol{w}\|$
\EndFor
\State \Return $L = L^{(T)}$
\end{algorithmic}
\end{algorithm}

In practice, only a small number of iterations (e.g., $T \leq 20$) is sufficient to obtain an accurate estimate of the largest eigenvalue, 
which is then used to set the step size $\alpha = h/L$ in Algorithm~\ref{alg:dose_weighted_PnP_reconstruction}.

\section{Additional PPO Training Details}
\label{appendix:ppo}

The training loop in Algorithm~\ref{alg:ppo-ct} (main text) summarizes the high-level workflow of PPO in the CT environment.  
For clarity and readability, minibatch updates, entropy regularization, and advantage estimation are omitted in the main algorithm but are detailed here for reproducibility.  
We denote the parametric approximation of the value function by $V(\cdot;\boldsymbol{w}_v)$ to distinguish it from the (unknown) true value function $V^\pi(\cdot)$. This convention follows our previous work~\cite{wang2024sequential}.

\paragraph{Notation.}
\begin{itemize}
    \item $\widehat{\boldsymbol{x}}_k$: reconstructed state at step $k$,
    \item $\theta_k$: action (e.g., projection angle or dose) at step $k$,
    \item $r_k$: immediate reward (see Equation~\ref{eq:rewardfunction}),
    \item $\pi_a(\cdot \mid \widehat{\boldsymbol{x}}_k;\boldsymbol{w}_a)$: stochastic policy with parameters $\boldsymbol{w}_a$,
    \item $V(\widehat{\boldsymbol{x}}_k;\boldsymbol{w}_v)$: approximate value function (critic) with parameters $\boldsymbol{w}_v$,
    \item $G_k$: Monte Carlo return from step $k$,
    \item $\widehat{A}_k$: estimated advantage at step $k$,
    \item $e_k(\boldsymbol{w}_a) = \dfrac{\pi_a(\theta_k \mid \widehat{\boldsymbol{x}}_k;\boldsymbol{w}_a)}{\pi_a^{\text{old}}(\theta_k \mid \widehat{\boldsymbol{x}}_k)}$: policy probability ratio.
\end{itemize}

\paragraph{Minibatch Updates.}
As in standard PPO, parameter updates are performed over multiple epochs using minibatch sampling from the collected trajectory set. 
Although each episode produces only a small number of transitions ($M$ steps), we construct an effective training batch by aggregating transitions from multiple episodes until a predefined batch size (e.g., 64 samples) is reached. 
This ``batch-size accumulation'' strategy stabilizes optimization, ensures sufficient gradient diversity, and enables PPO updates even in environments where each trajectory is short (as in sparse-view CT).

\paragraph{Advantage Estimation.}
We use generalized advantage estimation (GAE) with discount factor $\gamma$ and parameter $\lambda \in [0,1]$.
Following the PPO update scheme, the temporal-difference residuals are computed using 
the value function parameters \(\boldsymbol{w}_v^{\text{old}}\), which are frozen during the update epoch:
\[
\delta_k = r_k 
+ \gamma V(\widehat{\boldsymbol{x}}_{k+1};\boldsymbol{w}_v^{\text{old}})
- V(\widehat{\boldsymbol{x}}_{k};\boldsymbol{w}_v^{\text{old}}).
\]
The advantages are then estimated as
\[
\widehat{A}_k 
= \sum_{t=0}^{T-k-1} (\gamma\lambda)^t\,\delta_{k+t},
\qquad 
G_k = \widehat{A}_k 
+ V(\widehat{\boldsymbol{x}}_{k};\boldsymbol{w}_v^{\text{old}}).
\]

\paragraph{Entropy Regularization.}
To encourage exploration, we add the policy entropy:
\[
\mathcal{H}\!\left(\pi_a(\cdot \mid \widehat{\boldsymbol{x}}_k)\right)
= - \sum_{\theta} \pi_a(\theta \mid \widehat{\boldsymbol{x}}_k)\,\log \pi_a(\theta \mid \widehat{\boldsymbol{x}}_k).
\]

\paragraph{Total PPO Objective.}
Combining the actor and critic losses from the main text with entropy regularization, the overall objective is
\[
J(\boldsymbol{w}_a,\boldsymbol{w}_v) =
\mathbb{E}\!\left[
    L^{\text{actor}}(\boldsymbol{w}_a)
    - c_v\,L^{\text{critic}}(\boldsymbol{w}_v)
    + c_e\,\mathcal{H}\!\left(\pi_a(\cdot\mid\widehat{\boldsymbol{x}}; \boldsymbol{w}_a)\right)
\right],
\]
where $c_v$ and $c_e$ are positive coefficients weighting the critic and entropy terms, respectively.

\bibliographystyle{IEEEtran}
\bibliography{reference}

@book{hansen2021computed,
  title={Computed tomography: algorithms, insight, and just enough theory},
  author={Hansen, Per Christian and J{\o}rgensen, Jakob and Lionheart, William RB},
  year={2021},
  publisher={SIAM}
}

@article{kazantsev1991information,
  title={Information content of projections},
  author={Kazantsev, IG},
  journal={Inverse problems},
  volume={7},
  number={6},
  pages={887},
  year={1991},
  publisher={IOP Publishing}
}

@article{ruthotto2018optimal,
  title={Optimal experimental design for inverse problems with state constraints},
  author={Ruthotto, Lars and Chung, Julianne and Chung, Matthias},
  journal={SIAM Journal on Scientific Computing},
  volume={40},
  number={4},
  pages={B1080--B1100},
  year={2018},
  publisher={SIAM}
}

@book{lindley1972bayesian,
  title={Bayesian statistics: A review},
  author={Lindley, Dennis Victor},
  year={1972},
  publisher={SIAM}
}

@article{ryan2016fully,
  title={Fully Bayesian optimal experimental design: A review},
  author={Ryan, EG and Drovandi, CC and McGree, JM and Pettitt, AN},
  journal={International Statistical Review},
  volume={84},
  pages={128--154},
  year={2016}
}

@article{rainforth2024modern,
  title={Modern Bayesian experimental design},
  author={Rainforth, Tom and Foster, Adam and Ivanova, Desi R and Bickford Smith, Freddie},
  journal={Statistical Science},
  volume={39},
  number={1},
  pages={100--114},
  year={2024},
  publisher={Institute of Mathematical Statistics}
}

@article{helin2022edge,
  title={Edge-promoting adaptive Bayesian experimental design for {X}-ray imaging},
  author={Helin, Tapio and Hyv{\"o}nen, Nuutti and Puska, Juha-Pekka},
  journal={SIAM Journal on Scientific Computing},
  volume={44},
  number={3},
  pages={B506--B530},
  year={2022},
  publisher={SIAM}
}

@article{barbano2022bayesian,
  title={Bayesian experimental design for computed tomography with the linearised deep image prior},
  author={Barbano, Riccardo and Leuschner, Johannes and Antor{\'a}n, Javier and Jin, Bangti and Hern{\'a}ndez-Lobato, Jos{\'e} Miguel},
  journal={arXiv preprint arXiv:2207.05714},
  year={2022}
}

@article{batenburg2013dynamic,
  title={Dynamic angle selection in binary tomography},
  author={Batenburg, K Joost and Palenstijn, Willem Jan and Bal{\'a}zs, P{\'e}ter and Sijbers, Jan},
  journal={Computer Vision and Image Understanding},
  volume={117},
  number={4},
  pages={306--318},
  year={2013},
  publisher={Elsevier}
}

@article{dabravolski2014dynamic,
  title={Dynamic angle selection in {X}-ray computed tomography},
  author={Dabravolski, Andrei and Batenburg, Kees Joost and Sijbers, Jan},
  journal={Nuclear Instruments and Methods in Physics Research Section B: Beam Interactions with Materials and Atoms},
  volume={324},
  pages={17--24},
  year={2014},
  publisher={Elsevier}
}

@inproceedings{batenburg2011bounds,
  title={Bounds on the difference between reconstructions in binary tomography},
  author={Batenburg, K Joost and Fortes, Wagner and Hajdu, Lajos and Tijdeman, Robert},
  booktitle={International Conference on Discrete Geometry for Computer Imagery},
  pages={369--380},
  year={2011},
  organization={Springer}
}

@inproceedings{elata2025adaptive,
  title={Adaptive compressed sensing with diffusion-based posterior sampling},
  author={Elata, Noam and Michaeli, Tomer and Elad, Michael},
  booktitle={European Conference on Computer Vision},
  pages={290--308},
  year={2025},
  organization={Springer}
}

@article{shen2022learning,
  title={Learning to scan: A deep reinforcement learning approach for personalized scanning in {CT} imaging},
  author={Shen, Ziju and Wang, Yufei and Wu, Dufan and Yang, Xu and Dong, Bin},
  journal={Inverse Problems \& Imaging},
  volume={16},
  number={1},
  year={2022}
}

@article{wang2024sequential,
  author={Wang, Tianyuan and Lucka, Felix and van Leeuwen, Tristan},
  journal={IEEE Transactions on Computational Imaging}, 
  title={Sequential Experimental Design for X-Ray CT Using Deep Reinforcement Learning}, 
  year={2024},
  volume={10},
  number={},
  pages={953-968},
  doi={10.1109/TCI.2024.3414273}}

@article{wang2024task,
  title={Task-Adaptive Angle Selection for Computed Tomography-Based Defect Detection},
  author={Wang, Tianyuan and Florian, Virginia and Schielein, Richard and Kretzer, Christian and Kasperl, Stefan and Lucka, Felix and van Leeuwen, Tristan},
  journal={Journal of Imaging},
  volume={10},
  number={9},
  pages={208},
  year={2024}
}

@article{wang2025dynamic,
  title={Dynamic Angle Selection in X-Ray CT: A Reinforcement Learning Approach to Optimal Stopping},
  author={Wang, Tianyuan and Lucka, Felix and Pelt, Dani{\"e}l M and Batenburg, K Joost and van Leeuwen, Tristan},
  journal={arXiv preprint arXiv:2503.12688},
  year={2025}
}

@article{shen2023bayesian,
  title={Bayesian sequential optimal experimental design for nonlinear models using policy gradient reinforcement learning},
  author={Shen, Wanggang and Huan, Xun},
  journal={Computer Methods in Applied Mechanics and Engineering},
  volume={416},
  pages={116304},
  year={2023},
  publisher={Elsevier}
}

@book{sutton2018reinforcement,
  title={Reinforcement learning: An introduction},
  author={Sutton, Richard S and Barto, Andrew G},
  year={2018},
  publisher={MIT press}
}

@article{schulman2017proximal,
  title={Proximal policy optimization algorithms},
  author={Schulman, John and Wolski, Filip and Dhariwal, Prafulla and Radford, Alec and Klimov, Oleg},
  journal={arXiv preprint arXiv:1707.06347},
  year={2017}
}

@article{wang2006penalized,
  title={Penalized weighted least-squares approach to sinogram noise reduction and image reconstruction for low-dose X-ray computed tomography},
  author={Wang, Jing and Li, Tianfang and Lu, Hongbing and Liang, Zhengrong},
  journal={IEEE transactions on medical imaging},
  volume={25},
  number={10},
  pages={1272--1283},
  year={2006},
  publisher={IEEE}
}

@article{fessler1994penalized,
  title={Penalized weighted least-squares image reconstruction for positron emission tomography},
  journal={IEEE transactions on medical imaging},
  volume={13},
  number={2},
  pages={290--300},
  year={1994},
  publisher={IEEE}
}

@article{xia2021ct,
  title={CT reconstruction with PDF: Parameter-dependent framework for data from multiple geometries and dose levels},
  author={Xia, Wenjun and Lu, Zexin and Huang, Yongqiang and Liu, Yan and Chen, Hu and Zhou, Jiliu and Zhang, Yi},
  journal={IEEE Transactions on Medical Imaging},
  volume={40},
  number={11},
  pages={3065--3076},
  year={2021},
  publisher={IEEE}
}

@article{schulman2015high,
  title={High-dimensional continuous control using generalized advantage estimation},
  author={Schulman, John and Moritz, Philipp and Levine, Sergey and Jordan, Michael and Abbeel, Pieter},
  journal={arXiv preprint arXiv:1506.02438},
  year={2015}
}

@article{rudin1992nonlinear,
  title={Nonlinear total variation based noise removal algorithms},
  author={Rudin, Leonid I and Osher, Stanley and Fatemi, Emad},
  journal={Physica D: nonlinear phenomena},
  volume={60},
  number={1-4},
  pages={259--268},
  year={1992},
  publisher={Elsevier}
}

@article{chambolle2004algorithm,
  title={An algorithm for total variation minimization and applications},
  author={Chambolle, Antonin},
  journal={Journal of Mathematical imaging and vision},
  volume={20},
  number={1},
  pages={89--97},
  year={2004},
  publisher={Springer}
}

@inproceedings{venkatakrishnan2013plug,
  title={Plug-and-play priors for model based reconstruction},
  author={Venkatakrishnan, Singanallur V and Bouman, Charles A and Wohlberg, Brendt},
  booktitle={2013 IEEE global conference on signal and information processing},
  pages={945--948},
  year={2013},
  organization={IEEE}
}

@inproceedings{kohler2004projection,
  title={A projection access scheme for iterative reconstruction based on the golden section},
  author={Kohler, Thomas},
  booktitle={IEEE Symposium Conference Record Nuclear Science 2004.},
  volume={6},
  pages={3961--3965},
  year={2004},
  organization={IEEE}
}

@article{craig2023real,
  title={Real-time tilt undersampling optimization during electron tomography of beam sensitive samples using golden ratio scanning and RECAST3D},
  author={Craig, Timothy M and Kadu, Ajinkya A and Batenburg, Kees Joost and Bals, Sara},
  journal={Nanoscale},
  volume={15},
  number={11},
  pages={5391--5402},
  year={2023},
  publisher={Royal Society of Chemistry}
}

@article{hendriksen2021tomosipo,
  title={Tomosipo: fast, flexible, and convenient 3D tomography for complex scanning geometries in Python},
  author={Hendriksen, Allard A and Schut, Dirk and Palenstijn, Willem Jan and Vigan{\'o}, Nicola and Kim, Jisoo and Pelt, Dani{\"e}l M and Van Leeuwen, Tristan and Joost Batenburg, Kees},
  journal={Optics Express},
  volume={29},
  number={24},
  pages={40494--40513},
  year={2021},
  publisher={Optical Society of America}
}

@article{kingma2014adam,
  title={Adam: A method for stochastic optimization},
  author={Kingma, Diederik P and Ba, Jimmy},
  journal={arXiv preprint arXiv:1412.6980},
  year={2014}
}

@article{andriiashen2024x,
  title   = {X-ray Image Generation as a Method of Performance Prediction for Real-Time Inspection: A Case Study},
  author  = {Andriiashen, Vladyslav and van Liere, Robert and van Leeuwen, Tristan and Batenburg, K. Joost},
  journal = {Journal of Nondestructive Evaluation},
  volume  = {43},
  number  = {3},
  pages   = {79},
  year    = {2024}
}

@article{graas2025scintillator,
  title={Scintillator decorrelation for self-supervised x-ray radiograph denoising},
  author={Graas, Adriaan and Lucka, Felix},
  journal={Measurement Science and Technology},
  volume={36},
  number={6},
  pages={065415},
  year={2025},
  publisher={IOP Publishing}
}

@article{mccollough2006ct,
  title={CT dose reduction and dose management tools: overview of available options},
  author={McCollough, Cynthia H and Bruesewitz, Michael R and Kofler Jr, James M},
  journal={Radiographics},
  volume={26},
  number={2},
  pages={503--512},
  year={2006},
  publisher={Radiological Society of North America}
}

@article{pelt2022foam,
  title={Foam-like phantoms for comparing tomography algorithms},
  author={Pelt, Dani{\"e}l M and Hendriksen, Allard A and Batenburg, Kees Joost},
  journal={Synchrotron Radiation},
  volume={29},
  number={1},
  pages={254--265},
  year={2022},
  publisher={International Union of Crystallography}
}

@article{kazantsev2017novel,
  title={A novel tomographic reconstruction method based on the robust Student's t function for suppressing data outliers},
  author={Kazantsev, Daniil and Bleichrodt, Folkert and van Leeuwen, Tristan and Kaestner, Anders and Withers, Philip J and Batenburg, Kees Joost and Lee, Peter D},
  journal={IEEE Transactions on Computational Imaging},
  volume={3},
  number={4},
  pages={682--693},
  year={2017},
  publisher={IEEE}
}

\end{document}